\documentclass[twocolumn,tighten]{aastex631}

\usepackage{amsthm,amsmath,amssymb}
\usepackage{mathrsfs}
\newcommand{\unit}[1]{{\ \rm #1}}
\usepackage{multirow}

\usepackage{bm}
\usepackage{enumitem}
\newcommand{\sinc}{\rm{sinc }}
\usepackage{amsmath}
\usepackage{threeparttable}

\shorttitle{Constraining Galactic Structure Using GWs from DWDs}
\shortauthors{Zhang, Deng, Lu, \& Yu}

\begin{document}

\title{
Constraining the Galactic Structure using Time Domain Gravitational Wave Signal from Double White Dwarfs Detected by Space Gravitational Wave Detectors
}

\correspondingauthor{Youjun Lu}

\author[0009-0002-1814-6887]{Siqi Zhang}
\affiliation{National Astronomical Observatories, Chinese Academy of Sciences, 20A Datun Road, Beijing 100101, China}
\affiliation{School of Astronomy and Space Science, University of Chinese Academy of Sciences,
19A Yuquan Road, Beijing 100049, China}
\author[0000-0001-8075-0909]{Furen Deng}
\affiliation{National Astronomical Observatories, Chinese Academy of Sciences, 20A Datun Road, Beijing 100101, China}
\affiliation{School of Astronomy and Space Science, University of Chinese Academy of Sciences,
19A Yuquan Road, Beijing 100049, China}
%
%
\author[0000-0002-1310-4664]{Youjun Lu}\thanks{luyj@nao.cas.cn}
\affiliation{National Astronomical Observatories, Chinese Academy of Sciences, 20A Datun Road, Beijing 100101, China}
\affiliation{School of Astronomy and Space Science, University of Chinese Academy of Sciences,
19A Yuquan Road, Beijing 100049, China}
\author[0000-0003-4859-3231]{Shenghua Yu}
\affiliation{National Astronomical Observatories, Chinese Academy of Sciences, 20A Datun Road, Beijing 100101, China}
%

\begin{abstract}
The Gravitation Wave (GW) signals from a large number of double white dwarfs (DWDs) in the Galaxy are expected to be detected by space GW detectors, e.g., the Laser Interferometer Space Antenna (LISA), Taiji, and Tianqin in the millihertz band. In this paper, we present an alternative method by directly using the time-domain GW signal detected by space GW detectors to constrain the anisotropic structure of the Galaxy. The information of anisotropic distribution of DWDs is naturally encoded in the time-domain GW signal because of the variation of the detectors' directions and consequently the pattern functions due to their annual motion around the sun. The direct use of the time-domain GW signal enables simple calculations, such as utilizing an analytical method to assess the noise arising from the superposition of random phases of DWDs and using appropriate weights to improve the constraints. We investigate the possible constraints on the scale of the Galactic thin disk and bulge that may be obtained from LISA and Taiji by using this method with mock signals obtained from population synthesis models. We further show the different constraining capabilities of the low-frequency signal (foreground) and the high-frequency signal (resolvable-sources) via the Markov Chain Monte Carlo method, and find that the scale height and length of the Galactic thin disk and the scale radius of bulge can be constrained to a fractional accuracy of $\sim$ $30\%$, $30\%$, $40\%$ (or $20\%$, $10\%$, $40\%$)
by using the low-frequency (or high-frequency) signal detected by LISA or Taiji.
\end{abstract}


\section{Introduction}
\label{sec:intro}

Space gravitational wave (GW) detectors, such as Laser Interferometer Space Antenna (LISA) \citep[LISA;][]{LISA2017}, Taiji \citep{Taiji2018}, and Tianqin \citep{Tianqin2016}, are expected to detect the GW signals in the millihertz (mHz) band. The target sources include the merging massive binary black holes ($\sim 10^4-10^7\unit{M_\odot}$), extreme/intermediate mass ratio inspirals, inspiralling stellar-mass binary black holes and neutron stars, Galactic double white dwarfs (DWDs), as well as the stochastic foreground and background due to the large amount of unresolved DWDs in the Galaxy and stellar mass compact binaries in extragalactic galaxies \citep[e.g.,][]{Cornish2007, Klein2016, Babak2017, Moore2017, Bonetti2020, Babak2023, Amaro2022, Lau2020, Lamberts2018, LISA2017, Amaro2023, Amaro2013, Amaro2007, Thorpe2019}. Among these sources, the DWDs may be the most common ones and about $10,000-60,000$ Galactic DWDs are expected to be detected individually by LISA, Taiji, and Tianqin \citep[e.g.,][]{Nelemans2004, Ruiter2010, Marsh2011, Yu2010, Korol2017, Korol2019, Korol2022, Huang2020, Kang2021, Zhang2022}, which may reveal abundant information about the physical properties, the formation and evolution processes of DWDs \citep[e.g.,][]{Lamberts2019, Li2020, Georgousi2023, Wilhelm2021, Adams2012, Edlund2005}. The rest ($\sim10^7-10^8$) DWDs, the majority of the DWD population in the LISA/Taiji band, cannot be resolved, but the GW signals from these sources combine together to form a stochastic foreground, which can be detected. The population properties may also be extracted from these DWD signal, one of which is the anisotropic distribution of the DWDs, an independent tracer of the anisotropic Galactic structure.  

The anisotropic distribution of those DWDs is encoded in the GW signal because of the direction change of the space GW detector constellation due to their annual motion around sun over the long observation period (four year or longer) and thus the annual variation of the response functions, which may be extracted from the GW signal. It has been investigated in the literature that the scale height of the Galactic disk may be well constrained by analysis using the spatial distribution of the resolved individual DWDs (extract from the observations) and the GW foreground in the frequency-domain \citep[e.g.,][]{Benacquista2006, Breivik2020,Korol2019,Lamberts2019, Li2020, Georgousi2023, Wilhelm2021, Adams2012, Gao2023}. 
In addition, a number of studies investigate the modulation of the mHz GW signals from the Galactic DWDs based on the time-domain GW signals or the time-frequency domain techniques, in which various properties of the GW signal (but not the Galaxy shape) were proposed to be constrained by the modulation profile \citep{Digman2022, Buscicchio2024, Criswell2024, Pozzoli2024}.
Here we present an alternative method by directly using the GW signal in the time-domain modulated by the annual motion of the detectors around sun to constrain the anisotropic structure of the Galaxy, which is simple and efficient. Furthermore, we explore the different constrain capability for low-frequency and high-frequency DWDs with our techniques.
By using mock GW signals from a population synthesis model, we demonstrate the feasibility of such a method for LISA and Taiji on constraining the scale height and length of the Galactic thin disk, and the scale radius of the bulge.

The paper is organized as follows. In Section~\ref{sec:sample}, we briefly describe how the mock DWD population in the Galaxy with disk and bulge components is obtained by population synthesis model. In Section~\ref{sec:method}, we show the calculation process of the GW signal from the mock Galactic DWDs. The noise analysis on parameter estimation and the signal and noise spectrum are detailed in this section. The detailed method to estimate the Galactic structure parameters by using the mock GW signal detected by Taiji and LISA with either a top-hat filter or an enhanced filter can also be seen in this section. We demonstrate the method validity via the Markov Chain Monte Carlo (MCMC) method and present our main results in Section~\ref{sec:result}. Conclusions and discussions are given in Section~\ref{sec:dis&con}.

\section{Mock population of the Galactic DWDs} 
\label{sec:sample}

In this work, we adopt an updated version of the population synthesis model in \citet{Yu2010} to mock the Galactic population of DWDs. In the model, the template bank for DWDs formed from binary stars is generated from the binary star evolution (BSE) code based on that described in \citet{Hurley2000} and \citet{Hurley2002} (see also \citet{Han1998}). According to the calculation, most DWDs are generated from two different channels, i.e., the Roche lobe overflow (RLOF) + common envelope (CE) ejection channel and the CE ejection + CE ejection channel, which produce $30$\% and $67$\% of the DWDs, respectively. With the DWD templates, we obtain both the geometric distribution of the DWDs and the DWD property distributions by assuming the star formation history of binary stars in different components of the Galaxy. We adopt a Galactic model which includes the components of bulge, thin disk, thick disk, and stellar halo. The DWDs distribute differently in these different Galactic components according to the calculation. The contribution of the stellar halo and the thick disk to the GW signal is small at the LISA/Taiji/Tianqin band. Therefore, in this paper we only consider the GW signal from those DWDs in the bulge and the thin disk for simplicity.

The stellar density distribution for these two components are set as follows. The bulge is assumed to be spherical and follows a stellar density distribution as
\begin{equation}
\rho_{\rm b}(r) = \frac{M_{\rm b}}{4\pi h_r^3}\exp\left[-\left(\frac{r}{h_r}\right)^2\right] M_\odot {\rm pc}^{-3},
\label{eq:bulge}
\end{equation}
where $r$ is the distance to the Galactic center, $h_r=0.5$\,kpc is the bulge scale length, and $M_{\rm b}=2\times10^{10}M_\odot$ is the total mass of the bulge. We also set a cutoff radius of $3.5$\,kpc for the bulge \citep{Nelemans2004}. The stellar density distribution of the thin disk is assumed to be described by
\begin{equation}
\rho_{\rm{td}}(R,z)=\frac{M_{\rm{td}}}{4\pi h_R^2h_z}\exp\left(-\frac{R}{h_R}\right){\rm sech}^2\left(-\frac{z}{h_z}\right) M_\odot {\rm pc}^{-3},
\label{eq:thindisk}
\end{equation}
where $R$ and $z$ are the natural cylindrical coordinates of the axisymmetric disk, $h_R=2.5$\,kpc is the scale length of the thin disk, $h_z=0.352$\,kpc is the scale height of the thin disk, and $M_{\rm{td}}=5.2\times10^{10}M_\odot$ is the total mass of the thin disk. The position of the Sun is set to $(R_\odot,z_\odot)=(8.5$\,kpc,$0.03)\,{\rm kpc}$ \citep{GRAVITY2019}.

We assume that the star formation rate in the bulge and thin disk is the combination of a minor star forming process (the first item of the following equation) and a main star formation process (the second item of the following equation) \citep{Yu2010, Smith1978}, and the total star formation rate is described as
\begin{eqnarray}
{\rm SFR}(t)&=&0.12((t-t_0)/\rm Gyr) \nonumber\\
&&+11\exp[-(t-t_0)/\tau] M_{\odot}{\rm yr}^{-1}, t>t_0,
\label{eq:SFR}
\end{eqnarray}
where $t$ is the elapse time, $t_0$ ($=4$\,Gyr) defines the formation time of the thin disk and bulge, and the current age of the bulge and thin disk is $10$\,Gyr, and $\tau=9$\,Gyr yields a current star formation rate in bulge and disk which are consistent with \citet{Smith1978, Timmes1997, Diehl2006}. After the formation of the DWDs, their orbits may decay and gradually circularize due to the GW radiation. However, the timescale for such orbital decay and circularization is long comparing with the Hubble time, and thus we ignore the orbital evolution after their formation.

\begin{figure*}
\centering
\includegraphics[width=0.7\textwidth]{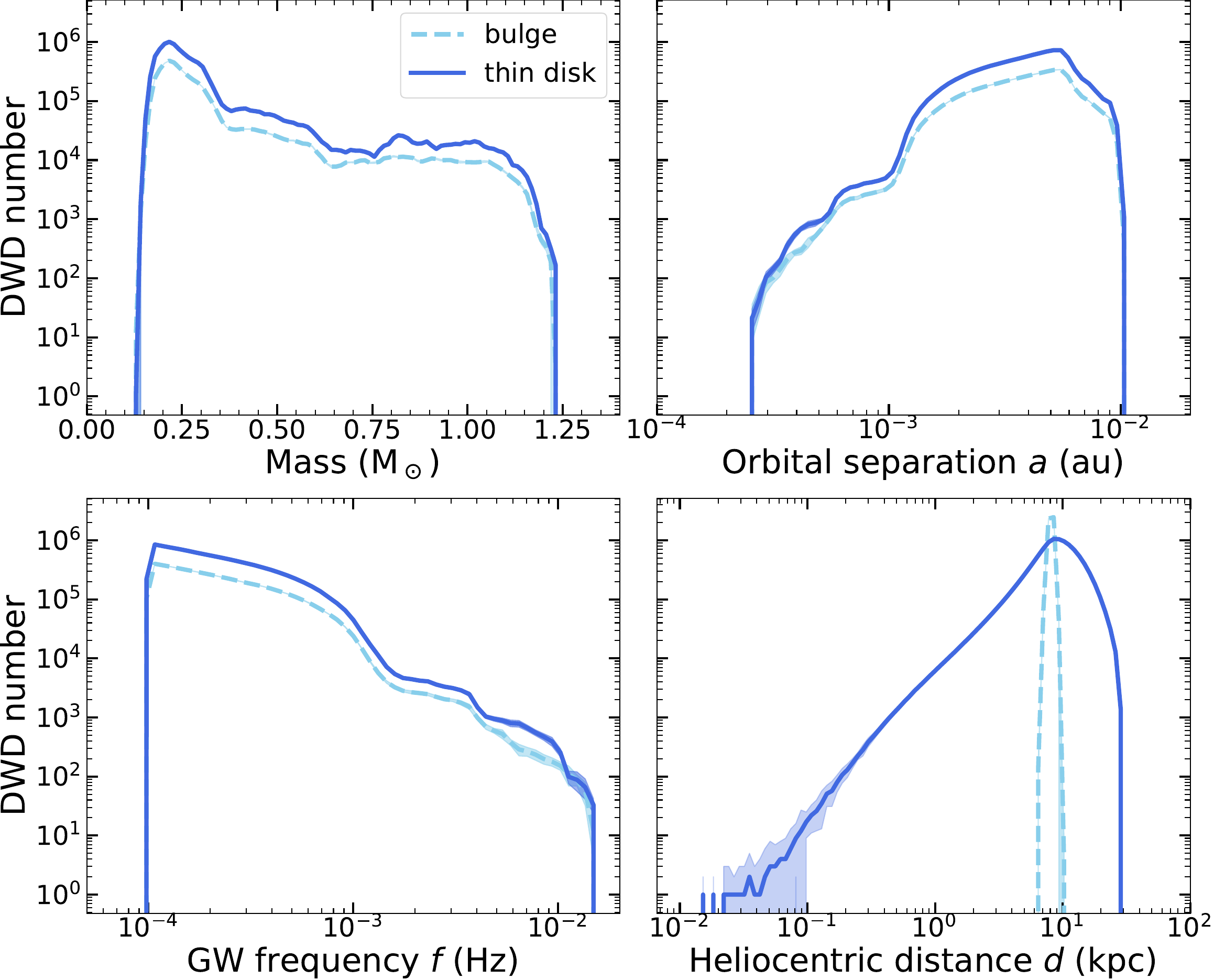}
\caption{
The chirp mass (top left panel), orbital separation (top right panel), GW frequency (bottom left panel), and heliocentric distance (bottom right panel) distributions of the mock Galactic DWDs. The solid blue and the dashed blue curves show the distribution of DWDs in the thin disk and the bulge, respectively. The mock Galactic DWD samples shown here are selected according to the frequencies of their GW radiation, i.e., $1\times10^{-4}\leq f \leq 1.5\times10^{-2}$\unit{Hz} (LISA and Taiji bands), and the total number of these mock DWDs in the thin disk and the bulge in this frequency range are $\sim 1.0\times 10^7$ and $4.9\times 10^6$, respectively. The lines and their associated shaded regions in each panel represent the mean values and the $3\sigma$ regions of the distributions, respectively, and the mean and standard error $\sigma$ in each bin are calculated from $10$ realizations generated from the population synthesis model for the Galactic DWDs. See details in Section~\ref{sec:sample}. 
}
\label{fig:DWDsamples}
\end{figure*}

By adopting the population synthesis approach outlined in \citet{Yu2010} and combining it with these density distributions, we are able to generate mock samples of DWDs in the Galaxy. To show the stochastic nature due to the limited number of DWDs, we generate totally $10$ different realizations of the Galactic DWDs. Figure~\ref{fig:DWDsamples} shows the distributions of four basic parameters of DWDs in mock sample(s), i.e., DWD chirp masses, orbital separations, GW frequencies, and heliocentric distances, with lines and shadows show the distributions of the mean distribution and the $3\sigma$ regions of the different realizations, respectively. As seen from this figure, the distributions of DWD parameters resulting from different realizations only have slight differences. The heliocentric distance distribution at small distances ($d\lesssim 2\times 10^{-2}$\,kpc) fluctuates the most due to the expected number of the DWDs within these distances is less than $100$ (bottom right panel), and the GW frequency distribution of the DWDs at the high-frequency end ($\gtrsim 10^{-2}$\,Hz; bottom left panel) also fluctuates as the number of the DWDs with orbital separation ($\lesssim 4\times 10^{-3}$\,AU) is  small ($\lesssim 10^2$; top right panel).

\section{Method} 
\label{sec:method}

In this section, we introduce our method to obtain constraints on the Galactic structure parameters by using the time-domain GW signals from DWDs detected by space GW detectors. We first obtain the mock GW signals from the Galactic DWDs in Section~\ref{sec:GWandSNR}-\ref{sec:signal spectrum}, and then consider the noises that may have in the signal in Section~\ref{sec:noise analysis}. A method to estimate the Galactic structure parameters by using mock GW signals with filters can be seen in Section~\ref{sec:filters}-\ref{sec:par-est}.

\subsection{GW signal from individual DWD} 
\label{sec:GWandSNR}

The GW radiation from an individual DWD can be considered as monochromatic as the orbital decay timescale of a DWD in the mHz band is much longer than the observation period of space GW detectors (i.e., $\nu\simeq$\,constant and $\frac{T_{\rm obs}}{\nu} \frac{d\nu}{dt}\rightarrow 0$ with $T_{\rm obs}$ denoting the observation period). The GW strain signal observed by a detector at any given time $t$ can be described as a combination of the $+$ and $\times$ polarizations, i.e.,
\begin{equation}
h(t)=\frac{\sqrt{3}}{2}\left(h_+F^++h_\times F^\times\right),
\end{equation}
where $\sqrt{3}/2$ is reduced from that the plane of the detector and the ecliptic plane is tilted by $60^\circ$. In the above Equation, 
\begin{equation}
h_+(t)=\mathcal{A}(1+\cos^2{\iota})\cos({2\pi \nu t + \phi_0 +\Phi_{\rm D}(t)}),
\label{hp}
\end{equation}
and
\begin{equation}
h_\times(t)=2\mathcal{A}\cos{\iota}\sin({2\pi \nu t + \phi_0 +\Phi_{\rm D}(t)}),
\label{hc}
\end{equation}
are the two polarizations, with 
$\mathcal{A}=\frac{2(G\mathcal{M})^{5/3}}{c^4d}(\pi \nu)^{2/3}$
($G$ and $c$ representing the gravitational constant and the speed of light, $d$ the luminosity distance of the source), and $\mathcal{M} = (m_1m_2)^{3/5}/(m_1+m_2)^{1/5}$ with $m_1$ and $m_2$ denoting the masses of its two components, respectively, $\iota$, $\Phi_{\rm D}(t)$, and $\phi_0$ represent the chirp mass, the inclination angle between the DWD orbital angular momentum and the vector pointing from the detector to the DWD, and the Doppler phase correction caused by the orbital motion of the space detectors around the sun, and the initial phase, respectively. The term $\Phi_{\rm D}(t)$ is given by \citep{cutler1998}
\begin{equation}
\Phi_{\rm{D}}(t) = 2\pi \nu \frac{R}{c}\sin\left(\frac{\pi}{2}-\beta\right)\cos\left(2\pi \nu_{\rm m} t-\lambda\right),
\label{eq:phi_D}
\end{equation}
where $R=1$\,AU is the distance from the earth to the sun, and $\nu_{\rm m} = 1$/year is the modulation frequency, $(\lambda, \beta)$ are the sky position in ecliptic coordinates of the source. The quantities $F^+$ and $F^\times$ represent the ``detector beam-pattern'' coefficients given by
\begin{eqnarray}
F^+(t,\theta_{\rm s},\phi_{\rm s},\psi_{\rm s}) & = &
\frac{1}{2}(1+\cos^2{\theta_{\rm s}})\cos{2\phi_{\rm s}(t)}\cos{2\psi_{\rm s}(t)} \nonumber \\
& & -\cos{\theta_{\rm s}}\sin{2\phi_{\rm s}(t)\sin{2\psi_{\rm s}(t)}},
\end{eqnarray}
and
\begin{eqnarray}
F^\times(t,\theta_{\rm s},\phi_{\rm s},\psi_{\rm s})& = & \frac{1}{2}(1+\cos^2{\theta_{\rm s}})\cos{2\phi_{\rm s}(t)}\sin{2\psi_{\rm s}(t)} \nonumber \\
& & +\cos{\theta_{\rm s}}\sin{2\phi_{\rm s}(t)\cos{2\psi_{\rm s}(t)}},
\end{eqnarray}
where $\theta_{\rm s}$ and $\phi_{\rm s}$ represent the position of the source in the detector's coordinate frame. The transformation between the detector coordinates and the ecliptic coordinates can be referred to \citet{cutler1998} and \citet{Wu2023} for LISA and Taiji, respectively.

According to \citet{Finn1996}, the signal-to-noise ratio (S/N) $\rho$ of a GW signal is given as
\begin{equation}
\rho^2=4\int^{\infty}_{0}\frac{\tilde{h}(\nu)\tilde{h}^*(\nu)}{S_{\rm n}(\nu)}d\nu,
\label{SNR}
\end{equation}
where $\tilde{h}(\nu)$ is the Fourier transform of the GW signal $h(t)$ and $S_{\rm n}(\nu)$ is the one-side power spectral density for the Gaussian noise of the detector. For a monochromatic signal, Equation~\eqref{SNR} can be re-written as 
\begin{equation}
\rho^2 = \frac{2}{S_{\rm n}(\nu)}\int^{T_{\rm obs}}_0 h^2(t)dt
\label{eq:SNR}
\end{equation}
using the Parseval's theorem with monochromatic frequency $\nu$ for each DWD source. One may calculate the S/N value for each given source by using the above formulae for a detector with $S_{\rm n}$. Given a typical S/N threshold, DWD sources can initially be divided into two categories, i.e., those above the S/N threshold and those below the threshold. In the detection frequency range of a space GW detector (e.g., $10^{-4}\sim10^{-1}$\,Hz for LISA or Taiji), even the S/N of some sources are above the threshold, they may be still stacked together and difficult to extract from the data, and only the DWDs with higher S/N can be resolved. However, the unresolved sources may be detected as the foreground signal. Note that there are also other GW sources in the LISA/Taiji/Tianqin band, such as binary stellar black holes and binary neutron stars, of which the GWs can also form a stochastic background. However, the GW background from these sources may be weaker and different from that from DWDs in the mHz frequency range \citep{Zhao2021, Pan2020, Adams2014, Bender1997,Xuan2024}, therefore, they are supposed to be ignored and only the GW signals from the Galactic DWDs should be considered.

It is computationally expensive to sample each of the disk DWDs over a long observation period due to the enormous number of DWDs in the sample. Given that the GW signal of each DWD is a sinusoidal function (see Eqs.~\ref{hp} and \ref{hc}), and considering the substantial difference between the frequencies of the sources ($\sim$\,mHz) and the detector motion ($\sim $ tens of nanohertz), it is feasible to firstly analytically compute the integral for sinusoidal term in Equation~\eqref{eq:SNR} for disk DWD signals in each time chunk and then numerically compute the detector's response which has long period \citep{Huang2020, Korol2017}. Therefore, it no longer needs expensive computation to sample the signal of disk DWDs to meet the requirements of the Nyquist sampling theorem. According to this, one can easily calculate the S/N for each source using the detected time-domain signal and select out those detectable DWDs with $\rho$ larger than a threshold S/N $\rho_{\rm thr}$. Subsequently, we obtain an irreducible GW foreground by subtracting the detectable or resolved sources \citep{Korol2022, Robson2017}. In this paper, we simply set $\rho_{\rm{thr}}=7$. However, the influence of varying S/N threshold on the results is discussed in Section~\ref{sec:result}.

\begin{figure}
\centering
\includegraphics[width=0.48\textwidth]{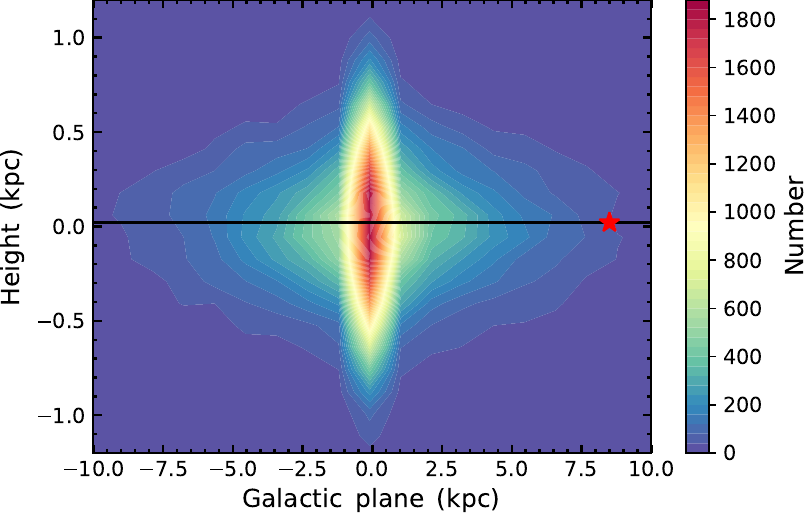}
\caption{
The 2-D projection of the spatial distribution of those resolvable DWDs detected by LISA (the total number is 43682). The x- and y-axes indicate the location of DWDs in the Galactocentric rest frame, i.e., the distance of the DWD projected onto the Galactic plane to the Galactic center and the height of the DWD above the Galactic plane, respectively. Different colors represent different number density of those resolvable DWDs as indicated by the right colorbar. The position of the sun $(8.5$\,kpc,$0.03$\,kpc) is marked by a red star. 
}
\label{fig:resolve_distribution}
\end{figure}

Figure~\ref{fig:resolve_distribution} shows the 2-D projection of the spatial distribution of the detectable/resolvable DWDs for illustration. As seen from this figure, the detectable/resolvable DWDs detected by GW space detector(s) trace the anisotropy of the Galaxy and thus can be used to constrain the Galactic structure parameters \citep[e.g.,][]{Korol2019, Lamberts2019, Li2020, Georgousi2023, Wilhelm2021, Adams2012, Gao2023}. The foreground formed by a large number ($\sim10^7-10^8$) of the un-resolvable DWDs can also be detected by LISA and Taiji. The anisotropic distribution of these DWDs can also be used to reveal the Galactic structure parameters \citep[e.g.,][]{Benacquista2006, Breivik2020}.

\subsection{Combined GW Signal from all Galactic DWDs}
\label{sec:signal_in_time_domain}

In this paper, we investigate the possible constraints on the Galactic structure that may be obtained by utilizing the GW signal of DWDs in the time domain. The information about the anisotropic distribution of DWDs in the Galaxy is encoded in the detected GW signals due to both the positional change of the detectors and the changes of the direction of the detector antennas caused by their rotation around the sun (and the motion of the detectors around the Galactic center). The positional change during the observation period of the detectors is much smaller than the anisotropy scale of the DWDs in the Galaxy, therefore the effect due to the changes of detector position can be ignored. The yearly directional changes of the detectors' (LISA/Taiji) antennas are significant and lead to significant changes in the response to the GW signals from different sky areas, of which the anisotropy is mainly determined by the geometry of different Galactic structures. Thus constraints on the Galactic structure may be extracted from the annual modulated time-domain signal. Taking into account the above, we consider extracting yearly modulate profile of GW signal in the time-domain.

We divide the square of the total time-domain signal into different data chunks to obtain the yearly modulated profile (hereafter as ``YMP'') of the GW signal, denoting each chunk by a subscript $k$, and then we can obtain the YMP profile by averaging the squared GW signal in the chunk over time as
\begin{equation}
F_k =  \frac{1}{\Delta{T}}\int^{(k+1)\Delta{T}}_{k\Delta{T}} h^2(t)dt,
\label{eq:Fk}
\end{equation}
where $h(t)$ is the detected time-domain signal for all DWD sources and $\Delta{T}$ is the length for each time chunk. The time duration $\Delta T$ may be set in the range from several hours to several days, which is longer than the typical period of DWDs (several hours) but substantially smaller than the motion period of the detectors around the sun (a year) and thus can reflect the modulation of the GW signal. In this paper, we simply set it as $1$\,day. 

According to the procedures to obtain YMP, it is expected that the yearly-varying profile of the GW signals has significant fluctuations. The fluctuation is mainly caused by the almost randomly distributed phases of DWD GW signals which cannot be eliminated by averaging. Such a stochastic fluctuation has little help for estimating the anisotropic of the Galaxy and should be treated as a noise term, which we therefore denote it as the phase noise. We present detailed analysis and discussions for this phase noise in Section~\ref{sec:noise analysis}. In general, the amplitude of the phase noise depends on the choice of $\Delta T$. Note that adopting a larger $\Delta T$ can lead to a smaller phase noise, however, it results into a smaller number of data points of the YMP of the GW signal for parameter estimation and thus leads to larger statistical errors. Considering these effects together, somewhat different choices of $\Delta T$ may affect the results little. Note also that the Doppler term in Equation~\eqref{eq:phi_D} is ignored in our calculation as the fractional changes due to this term on the YMP is smaller than $1\%$ when adopting the average length $\Delta T = 1$\,day. 

We generate mock DWDs and consequently mock signals for data analysis as real observation is not available, yet. We derive an analytic expression to calculate the smoothly part of YMP using the mock DWD samples given by
\begin{equation}
\left<F_k\right>=\frac{\sum_n {\left(A_n^2+B_n^2\right)}}{2},
\label{eq:<FK>}
\end{equation}
where $A_n$ and $B_n$ represent the constant coefficients for the $+$ and $\times$ polarization of the GW signal from each DWD among those in each time chunk, respectively, which are determined by $\mathcal{A}$, $\iota$, and the average value of the pattern function in the corresponding time chunk. The values of these quantities are approximately constants in a time chunk with small duration (for detailed description, see the Equation~\eqref{eq:A_ht} and the text below it in Appendix~\ref{sec:appendixA}). According to this equation, the mock smooth YMP profile can be easily obtained from the DWDs resulting from a Galactic model with any given set of structure parameters. With this smooth profile, the model term in parameter estimation can be obtained.

\begin{figure*}
\centering\includegraphics[width=1\textwidth]{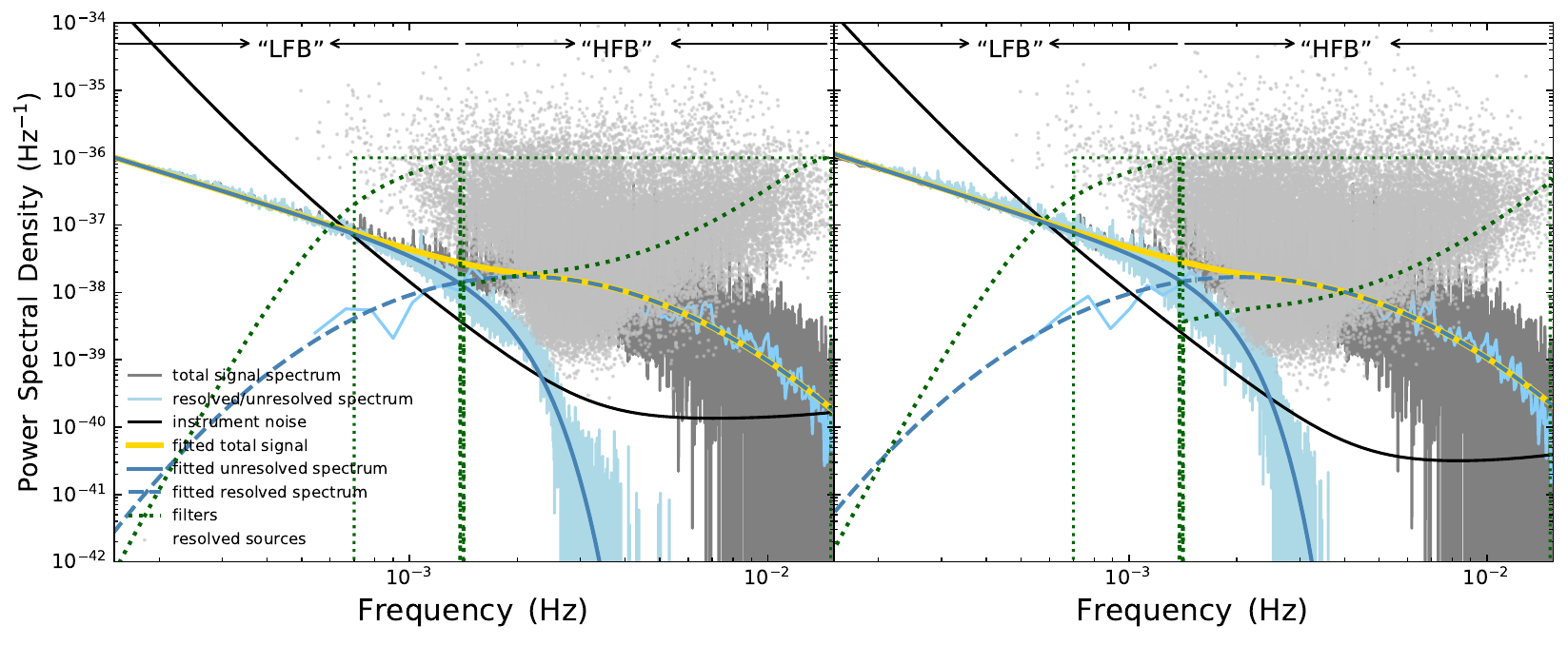}
\caption{
The power spectral density of the GW signal from the Galactic DWDs. For the left panel, the grey curve shows the total spectrum of the GW signal from all DWD sources and the two light blue curves which peaks at higher/lower frequency show the spectrum of the GW signals from all the resolvable DWDs/the unresolvable DWDs by LISA 4-year (left panel)  or Taiji 4-year (right panel) observations, respectively. These curves are smoothed with a window of $100$ frequency bins and the length of each bin is $3.037\times 10^{-8}$\,Hz. The solid dark blue curve and the dashed dark blue curve show the fitting results obtained for the spectra of the foreground (unresolvable) signal and the resolvable signals, respectively. The yellow curve shows the best fit to the total spectrum. The grey points represent the resolvable DWD sources after 4-year observations by LISA or Taiji. In each panel, four green dotted lines show different filters (see Section~\ref{sec:filters}). The lines with frequency window $7\times10^{-4}\sim1.4\times10^{-3}$\,Hz  and $1.4\times10^{-3}\sim1.5\times10^{-2}$\,Hz are the top-hat filters for the ``LFB'' and ``HFB'' bands, respectively. The line with a peak around $1.5\times10^{-3}$\,Hz is the enhanced filter for the ``LFB'' band and the line with a peak around $1.5\times10^{-2}$\,Hz is the enhanced filter for the ``HFB'' band. The maximum value for each filter is scaled for the purpose of demonstration, and the true maximum values are listed in Table~\ref{tab:amplitude for filters}.
}
\label{fig:spectrum_LISAandTaiji}
\end{figure*}

\begin{table}
\centering
\caption{The true maximum amplitude of each filter. }
\begin{tabular}{cccc} \hline \hline
Filter & $w_{\rm th}$ & $w_{\rm{op}}$ (``LFB'') & $w_{\rm{op}}$ (``HFB'')  \\ \hline
LISA & 1 & $3.880\times10^{37}$ & $3.010\times10^{39}$ \\ \hline
Taiji & 1 &$3.878\times10^{37}$ & $1.060\times10^{40}$ \\ \hline \hline
\end{tabular}
\label{tab:amplitude for filters}
\end{table}

\subsection{Signal spectrum}
\label{sec:signal spectrum}

We calculate the GW spectrum from mock Galactic DWDs to analyze the properties of these sources. The spectrum can be obtained by summing the square of each source amplitude as $\sum_n (A_n^2 + B_n^2) $ in any time chunk to obtain the shape of the spectrum $S_0(\nu)$ for one day. Then, the time-domain signal for 4-year observation can be used to obtain the normalization amplitude of the spectrum as  
\begin{equation}
A_k = \frac{\left<F_k\right>}{\int S_0(\nu) d\nu},
\label{eq:A_k}
\end{equation}
by using Parseval's theorem, and the final spectrum can then be written as
\begin{equation}
S_s = A_k S_0(\nu).
\label{eq:Ss}
\end{equation}
In the calculation of the GW spectrum, we assume that the shape of the spectrum changes little across time chunks. This assumption is reasonable. Assuming a sufficiently large number of sources in each direction and the properties of these sources are location independent, the only variations of the GW signals from different directions (or time chunks) are caused by two factors. One is that at different directions the number of sources are different and the other is the annual variation of the pattern function due to direction changes of the detectors' antennas. The former assumption is supported by the vast number of DWDs present in the Galaxy, exceeding $10^7$ (even for the detectable/resolvable sources, still exceeding tens of thousands). For the latter assumption, it is valid if we assume that the Galaxy is composed of a single component, and the properties of sources in this component are location independent. However, in our calculation, we consider two distinct components, the disk and the bulge. Nonetheless, according to the bottom left panel of Figure~\ref{fig:DWDsamples}, one can see that the difference between the spectrum by bulge and that by disk is small. With the fact that the responses of detectors' antennas are broad, it is reasonable to assume that the properties of sources change little across different directions (or time chunks) for these two components. Therefore, we can reasonably assume that the properties of sources in each direction are mostly identical and the final spectrum can be estimated by the shape calculated from any time chunk with one-day length times the amplitude of the spectrum which can be calculated using the total time-domain amplitude (see Eq.~\ref{eq:Ss}). 

According to Equations~\eqref{eq:A_k}-\eqref{eq:Ss}, we calculate the spectrum for all DWD sources ($S_s$) with averaged $\left<F_k\right>_{\rm{all}}$ as described in Section~\ref{sec:signal_in_time_domain}. The spectrum can be seen in Figure~\ref{fig:spectrum_LISAandTaiji} in grey for LISA and Taiji. Combining with the detector sensitivity curve, we can see that the GW signal spectrum of DWDs have different detection capability across different frequency range. Subsequently, we calculate both the foreground spectrum and the spectrum of resolved sources, with the aim of showing the relationship between detection capabilities and the respective frequency ranges. To obtain these two spectrum, we calculate the YMP $\left<F_k\right>_{\rm{fg}}$ and $\left<F_k\right>_{\rm{res}}$ for foreground and resolved sources, respectively. And then we use the same method as all source to get the foreground spectrum ($S_{\rm{ur}}$) and the spectrum from the resolved sources ($S_{\rm{r}}$). These spectrum are shown in Figure~\ref{fig:spectrum_LISAandTaiji} in light blue with solid and dashed lines, respectively. We can see that at lower frequency ($\leq 1.4\times10^{-3}\unit{Hz}$), the spectrum is dominated by foreground and at higher frequency ($\geq 1.4\times10^{-3}\unit{Hz}$), the spectrum is dominated by the spectrum of resolved DWD sources. To show the different constrain capability for different frequency range, we define two different ranges: one is called ``high frequency band'' (abbreviated as ``HFB'') and the other is ``low frequency band'' (abbreviated as ``LFB'') to show their different constrain capability of the Galactic structure in this work. It is worth to mention that, the amplitude of the spectrum changes among different time chunks, however for the demonstrate purpose, we show the average amplitude among different time chunk for them in this figure.
\subsection{Noise analysis}
\label{sec:noise analysis}

Estimating the noises in the observational data is important for parameter estimation using the YMP. The noises that affect the YMP should be considered, which include both the instrumental noise and the phase noise. The phase noise comes from the random phases of DWD GW signals in each time chunk. The detected signal comprises of thousands of sources and the phase of the GW signal from each DWD cannot be measured explicitly. Therefore, the fluctuation introduced by the (random) phases of the GW signals from different DWDs cannot be smoothed out and it can be taken as a kind of noise,  denoting as the phase noise, which is different from the instrumental noise.

Space GW detectors, such as LISA and Taiji, are considered in our calculations. The sensitivity curve for LISA can be well approximated by \citep{LISAcurve}
\begin{eqnarray}
S^{\rm{LISA}}_n(\nu) & = & \frac{10}{3L_{\rm{LISA}}^2}\left(P_{\rm{OMS}}(\nu)+\frac{4P_{\rm{acc}}(\nu)}{(2\pi \nu)^4}\right) \nonumber \\
& & \times \left(1+\frac{6}{10}\left(\frac{\nu}{\nu_*}\right)^4\right),
\label{SCforLISA}
\end{eqnarray}
where $L_{\rm{LISA}}=2.5\times10^9\unit{m}$ is the arm-length for LISA, $\nu_*=19.09\unit{mHz}$ is the transfer frequency \citep{LISAsc2}, $P_{\rm{OMS}}(\nu)$ and $P_{\rm{acc}}(\nu)$ are single-link optical metrology noise and single test mass acceleration noise, respectively, and their detail expressions can be consulted in \citet{LISAcurve}. For Taiji, the expression of sensitivity curve is similar with Equation~\eqref{SCforLISA} while the arm-length $L_{\rm{Taiji}}=3\times10^9\unit{m}$ and the $P_{\rm{OMS}}$ is about half of LISA \citep{Taiji,Liu2023}. The sensitivity curves for LISA and Taiji can be seen in Figures~\ref{fig:spectrum_LISAandTaiji} for illustration.

The YMP is calculated by squaring $h(t)$ and then averaging over time in each time chunk. Different realizations may result in somewhat different YMPs, and the variance of these YMPs in any given time chunk is not $0$ due to the random phases of DWDs. In principle, we can numerically realize the total signal many times, where each DWD contributing to the total signal has an independent random phase, and then calculate the variance in each time chunk, however, it is computation consuming. For simplicity, we derive an analytic expression to evaluated this variance and summarize it in Appendix~\ref{sec:appendixA}. Note that most DWDs in the LISA/Taiji band radiate monochromatic GW signals, thus there should be a correlation between signals at different times. However, as the proof we show in Appendix~\ref{sec:appendixA}, this correlation can be ignored if the number of sources is large enough in each time chunk and the duration of the chosen time chunk is large. These two assumptions are reasonable for the fact that the immense number of DWDs in the Galaxy and the chosen time chunk duration of one day is substantially longer than the GW period of each individual DWD. The total noise, which comprises of two terms, one arising from the random phases of the DWD sources (denoted as phase noise) and the other stemming from the detector noise, can be written as
\begin{equation}
\sigma^2_k=\frac{1}{\Delta T}\delta_{k'k}\int_0^{+\infty}\left(S_s^2(\nu)+S_{\rm n}^2(\nu)\right) d\nu , 
\label{sigma}
\end{equation}
where the $\delta_{k'k}$ is the Kronecker delta, it shows that the covariance between different time chunks can be ignored, $S_s$ is the power spectral density of the GW signals from all sources. This equation contains the fluctuation caused by signal and the instrumental noise. From Equation~\eqref{sigma}, we can see that if we have the spectrum of signal and the spectrum of detector noise, we can estimate the variance for the averaged-total signal which makes it possible for parameter estimation by using the profile of the time-domain GW signals of DWDs in the Galaxy. The instrumental noise is described in this section and the spectrum of $S_s$ can be seen in Section~\ref{sec:signal spectrum} and they are all shown in Figure~\ref{fig:spectrum_LISAandTaiji}. Therefore, with these spectrum, one can obtain the $\sigma_k$ by integrate the spectrum over the frequency band which we are interested in.

\begin{figure}
\centering
\includegraphics[width=0.5\textwidth]{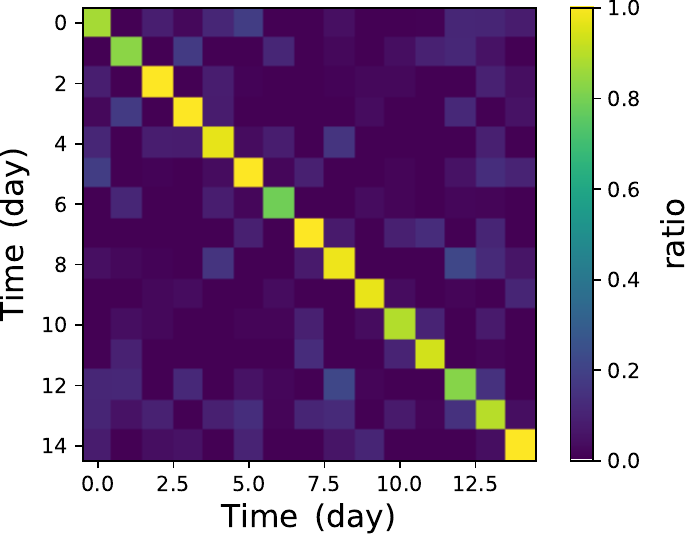}
\caption{
The covariance of the ratio of the standard deviation calculated from the analytic method to the standard deviation calculated form numerical method in $15$ consecutive time chunks. The duration of each time chunk is $1$\,day.
}
\label{fig:cov_sigma}
\end{figure}

To demonstrate the reasonableness of employing the analytic method for calculating $\sigma_k$, we compare the standard deviation calculated from the analytical method and the standard deviation calculated form the numerical method and show the covariance of their ratios in $15$ consecutive time chunks in Figure~\ref{fig:cov_sigma}. To obtain the numerical result, we calculate consecutive $15$ days of DWD signals detected by LISA and generate $150$ realizations of these signals with randomly initial phases for each source. Subsequently, we calculate the mean square value of the signal for each realization and compute the variance across these realizations for each time chunk ($1$\,day). As for the analytic result, it can be easily calculated based on the power spectral density by using Equation~\eqref{sigma}. We then use the ratio of the standard deviation obtained by these two methods to calculate the covariance. We can see that the diagonal terms are all close to unit and the non-diagonal terms are all close to zero, which validates the analytic expression for estimating $\sigma_k$ and suggests that the relationships between different time chunks can be safely ignored.

\begin{figure}
\centering
\includegraphics[width=0.48\textwidth]{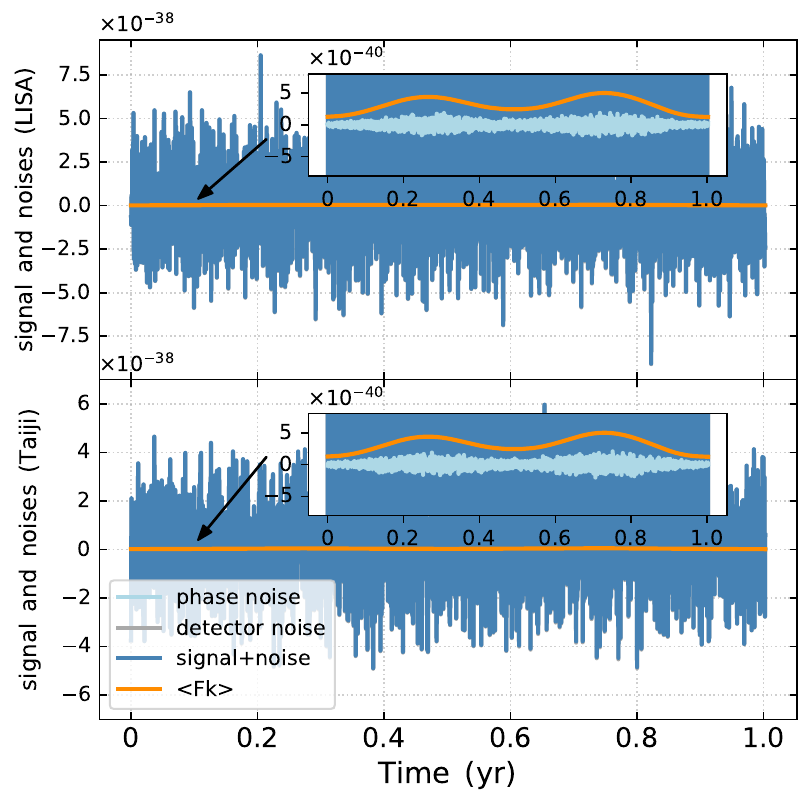}
\caption{
Time evolution of the curves (blue curves) by averaging the squared mock GW signal in each time chunk from the Galactic DWDs with inclusion of the standard deviation of the phase noise (cyan curves) and the detector noise (grey curves, which is covered by the blue curves in this figure due to the detector noise is much more larger than the signal in original situation) obtained from our Monte Carlo calculations for LISA (top panel) and Taiji (bottom panel). The orange curve represents the profile of mock time-domain averages signals without considering the detector noise and phase noise. The curves contain all DWD sources in LISA/Taiji sensitivity band $10^{-4}\sim1.5\times10^{-2}\unit{Hz}$. 
}
\label{fig:Fk_LISAandTaiji_original}
\end{figure}

Figure~\ref{fig:Fk_LISAandTaiji_original} shows the phase noise, detector noise (generated by assuming LISA or Taiji sensitivity), GW signal+noise and the profile of mock time-domain-average signals for LISA and Taiji. The signal shown in this figure contains all the DWD sources in the LISA/Taiji sensitivity band generated from the Galaxy model with structure parameters shown in Section~\ref{sec:sample}. Without any frequency selection or filtering applied, we can see that the detector noise is significantly higher than the GW signal, which may lead to poorer results of parameter estimations. This inspired us to take a series of measures to improve the constraining results in the subsequent calculations.

\subsection{Filters}
\label{sec:filters}

To place tighter constraints on the structural parameters of the Galaxy, we contemplate utilizing efficient filters.
We divide the GW signal into two different frequency bands, i.e., the ``HFB'' ($\geq 1.4\times10^{-3}\unit{Hz}$) and the ``LFB'' ($\leq 1.4\times10^{-3}\unit{Hz}$) in Section~\ref{sec:signal spectrum}, based on that the foreground is dominated at the ``LFB'' and resolvable DWDs are dominated at the ``HFB'' (see Fig.~\ref{fig:spectrum_LISAandTaiji}). To investigate the constraining power of the GW signals from these two different bands on the Galactic structure parameters, we first adopt top-hat filters to pick out the ``HFB'' and the ``LFB'' as
\begin{eqnarray}
w_{\rm{th}}(f)=
\begin{cases}
1, &\rm{if}\,\,\, \nu_{\rm{min}}\leq \nu \leq \nu_{\rm{max}},\\
0, &\rm{if}\,\,\, \nu< \nu_{\rm{min}} \,\,\rm{or}\,\, \nu>\nu_{\rm{max}},
\end{cases}
\label{eq:top-hat filter}
\end{eqnarray}
where the $\nu_{\rm{min}}$ and $\nu_{\rm{max}}$ denote for the minimum and maximum frequencies of a top-hat filter. We set $v_{\rm{min}} = 7\times10^{-4}$ and $v_{\rm{max}} = 1.4\times10^{-3}$ for ``LFB'', $v_{\rm{min}} = 1.4\times10^{-3}$ and $v_{\rm{max}} = 1.5\times10^{-2}$ for ``HFB'' in our calculation.

We may also consider enhanced filters, weighting the noise and the signal and thereby maximizing the S/N (both the noise caused by detector and phases), and thus may give better constraints on the Galactic structure parameters.
This enhanced filter can be written as (see Appendix~\ref{sec:appendixB} for the derivation)
\begin{equation}
w_{\rm{ehc}}(\nu)=\frac{S_s(\nu)}{S^2_s(\nu)+S_{\rm n}^2(\nu)},
\label{eq:op filter}
\end{equation}
where the $S_s$ is the spectrum for all sources, $S_{\rm n}$ is the sensitivity curve for LISA or Taiji. To show the difference in the constraining capability of the data in different frequency band (i.e., the different capability of ``LFB'' and ``HFB''), we also add a frequency truncation at $1.4\times10^{-3}\unit{Hz}$ to the enhanced filter.

Figure~\ref{fig:spectrum_LISAandTaiji} shows the enhanced and top-hat filters for LISA and Taiji in green dotted lines alongside the signal spectrum. For demonstration purpose, we scale the amplitude of these filters to $1\times10^{-36}$ as the maximum value, the true maximum values for these filters are summarized in Table~\ref{tab:amplitude for filters}. Note that the peak of the enhanced filter (``LFB'') is around $1.5\times10^{-3} \unit{Hz}$, corresponding to the frequency where the ratio of signal to detector noise is maximum. The peak of the optimal filters (``HFB'') is at the frequency around $1.5\times10^{-2} \unit{Hz}$, however, it is not around the frequency where the ratio of signal to detector noise is maximum. The reason for this difference between the enhanced filters for two frequency ranges is that the noise caused by random DWD phases dominates the total noise for the ``HFB'' but not for the ``LFB''. Therefore, the peak of the ``HFB'' is not at the frequency around the maximum ratio of signal and detector noise according to the definition of the enhanced filter in Equation~\eqref{eq:op filter}. 

With these two filters, the weighted signal and noise can be rewritten as
\begin{equation}
\sigma_w^2 = \frac{1}{\Delta T}\int_0^{+\infty}\left[w_0^2(\nu)\left(S_s^2(\nu)+S_{\rm n}^2(\nu)\right)\right]d\nu
\end{equation}
and
\begin{equation}
\langle F_w \rangle = \int_0^{+\infty} w_0(\nu) S_s(\nu) d\nu,
\end{equation}
where $w_0$ represents the top-hat filter $w_{\rm{th}}$ or the enhanced filter $w_{\rm{ehc}}$. By employing these filters, it is expected that the parameter estimation results will exhibit improved accuracy.

\begin{figure}
\centering
\includegraphics[width=0.49\textwidth]{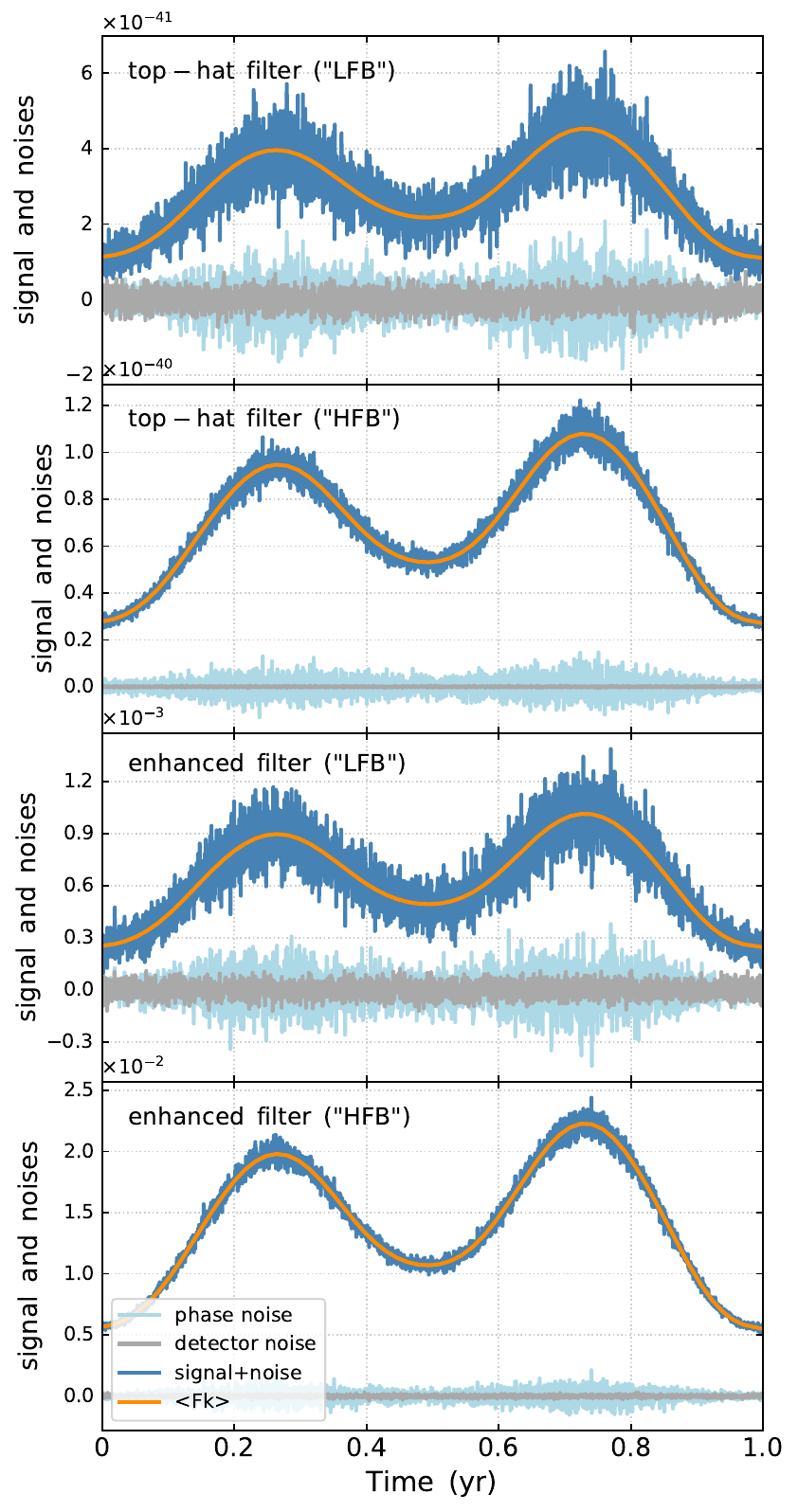}
\caption{
Time evolution of the curves (blue curves) by averaging the squared mock GW signal in each time chunk from the Galactic DWDs in either ``LFB'' or ``HFB'' with inclusion of the phase noise (cyan curves) obtained from our Monte Carlo calculations and the detector noise (grey curves, by assuming LISA observations). The orange curve represents the profile of mock time-domain averages signals without considering the detector noise and phase noise. From top to bottom, the panels show the results obtained by using the top-hat filter for ``LFB'' and ``HFB'', and the enhanced filter for ``LFB'' and ``HFB'', respectively.
}
\label{fig:Fk_LISA_filters}
\end{figure}

Figure~\ref{fig:Fk_LISA_filters} shows the standard deviation of the phase noise obtained from our Monte Carlo calculations, detector noise, signal+noise and the smoothly profile of mock time-domain average signals of LISA for different filters and frequency range. From top to bottom, the panels show the results obtained by using the top-hat filter to both ``LFB'' and ``HFB'', and the enhanced filter for ``LFB'' and ``HFB'', respectively. It can be seen that after applying filters, especially the enhanced filter, the noise can be suppressed well. Evidently, the noise level in the ``HFB'' is lower compared to the ``LFB'', attributed to the prevalent high S/N sources in the higher frequency range. 

\begin{figure*}
\centering
\includegraphics[width=1\textwidth]{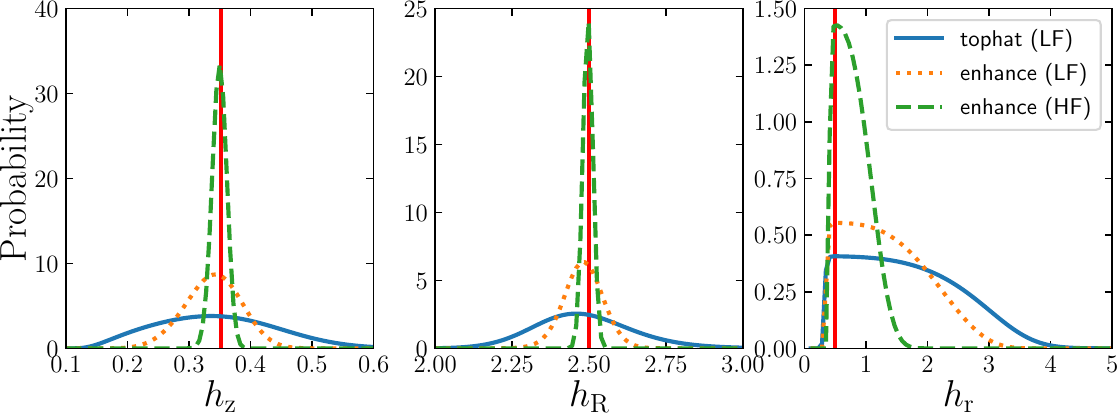}
\caption{
The marginalized probability distributions of the Galactic structure parameters obtained from the GW signal in different frequency band with different filters. Left to right panels show the results for $h_z$, $h_{\rm{R}}$, and $h_r$, respectively. In each panel, the solid line and the dotted line show the results obtained by using the ``LFB'' GW signals with the top-hat filter and enhanced filter, respectively; the dashed line shows the results obtained using the ``HFB'' GW signal with the enhanced filter; the vertical line indicates the input value of the Galactic structure parameter, i.e., $h_z=0.352$\,kpc, $h_{\rm{R}}=2.50$\,kpc, and $h_r=0.5$\,kpc.
}
\label{fig:marginal distribution}
\end{figure*}

\subsection{Parameter estimation method}
\label{sec:par-est}

The Bayesian fitting is the commonly used method to obtain the distribution of parameters \citep{Cutler1994, Grover2014, Veitch2015}. We adopt the Markov Chain Monte Carlo (MCMC) method to constrain the structure parameters of the Galaxy, mainly described by the scale height $h_z$ and scale length $h_R$ of the thin disk, and the scale radius $h_r$ of the bulge. The other two Galaxy components, i.e., the thick disk and the stellar halo, contribute little to the GW signal in the LISA/Taiji band \citep{Yu2010, Ruiter2009}, therefore, we ignore them in our calculations.

We assume a simple Gaussian likelihood as
\begin{eqnarray}   
\ln L(\bm{\theta}|\bm{\theta}_0) & = &  \sum_{k} \left( -\ln{\sqrt{2\pi}\sigma_k}+ \right. \nonumber \\
& & \left. \frac{1}{2\sigma_k^2}\left(\langle F_k(\bm{\theta})\rangle-\langle F_k(\bm{\theta}_0)\rangle\right)^2 \right),
\label{eq:likelihood}
\end{eqnarray}
where $\langle F_k\rangle$ is the time-domain yearly modulated profile in each time chunk (Eq.~\ref{eq:<FK>}), the subscript $k$ denotes the time chunk and $\bm{\theta}$ represents the structure parameter vector $\{h_z,h_R,h_r\}$ for the thin disk and the bulge and  $\bm{\theta}_0$ represents the true values of these parameters. The term $\langle F_k\rangle$ is composed of two parts, the bugle and the thin disk. To demonstrate our ability to constrain the amplitude of the signal, except for these three structure parameters, we also introduce two extra parameters $\chi_{\rm d}$ for the thin disk and $\chi_{\rm b}$ for the bulge as 
\begin{equation}
\langle F_k(\theta)\rangle = \chi_{\rm d}\langle F_k(\theta)\rangle_{\rm{disk}} + \chi_{\rm b}\langle F_k(\theta)\rangle_{\rm{bulge}},
\label{chi}
\end{equation}
which show the accuracy of the constraints on the amplitude of the YMP signal, with true values of $\chi_{\rm b}=\chi_{\rm d}=1$. Therefore, there are totally five parameters to be constrained by using the time domain signal profile, and we define the parameter vector as $\bm{\Theta}=\{h_z,h_R,h_r,\chi_{\rm d},\chi_{\rm b}\}$ instead of $\bm{\theta}$. Similarly, the true value of them can be written as $\bm{\Theta_0} = \{0.352$\,kpc, $2.5$\,kpc, $0.5$\,kpc, $1, 1\}$ instead of $\bm{\theta_0}$ in Equation~\eqref{eq:likelihood} \citep{Yu2010, Juric2018, Mackereth2017}.

Combining with the noise term, we can use the MCMC method by utilizing the likelihood described in Equation~\eqref{eq:likelihood} to obtain the posteriors of the model parameters.

\section{Results}
\label{sec:result}

In this section, we present the main results on constraining the structure parameters of the Galaxy by using the time-domain signal information.

Based on our model, we show the marginal probability distribution of three structure parameters by LISA in Figure~\ref{fig:marginal distribution}. We also display different situations in this figure, they are, applying top-hat and enhanced filters for the ``LFB'' and applying the enhanced filters for the ``HFB'', which are distinctively depicted using solid, dotted, and dashed lines, respectively. As seen from this figure, the marginalized probability distribution for the scale height $h_z$ shows smaller scatter compared to the other two parameters. Moreover, the ``HFB'' displays the smaller scatter than ``LFB''. For the distribution of $h_r$, it can be seen that the probability goes up faster when the value is small, this is because the value of scale radius should be positive, therefore we cut the minimum value of it at 0.1.

\subsection{Constraining capability on the structure parameters}

\begin{figure*}
\centering
\includegraphics[width=0.7\textwidth]{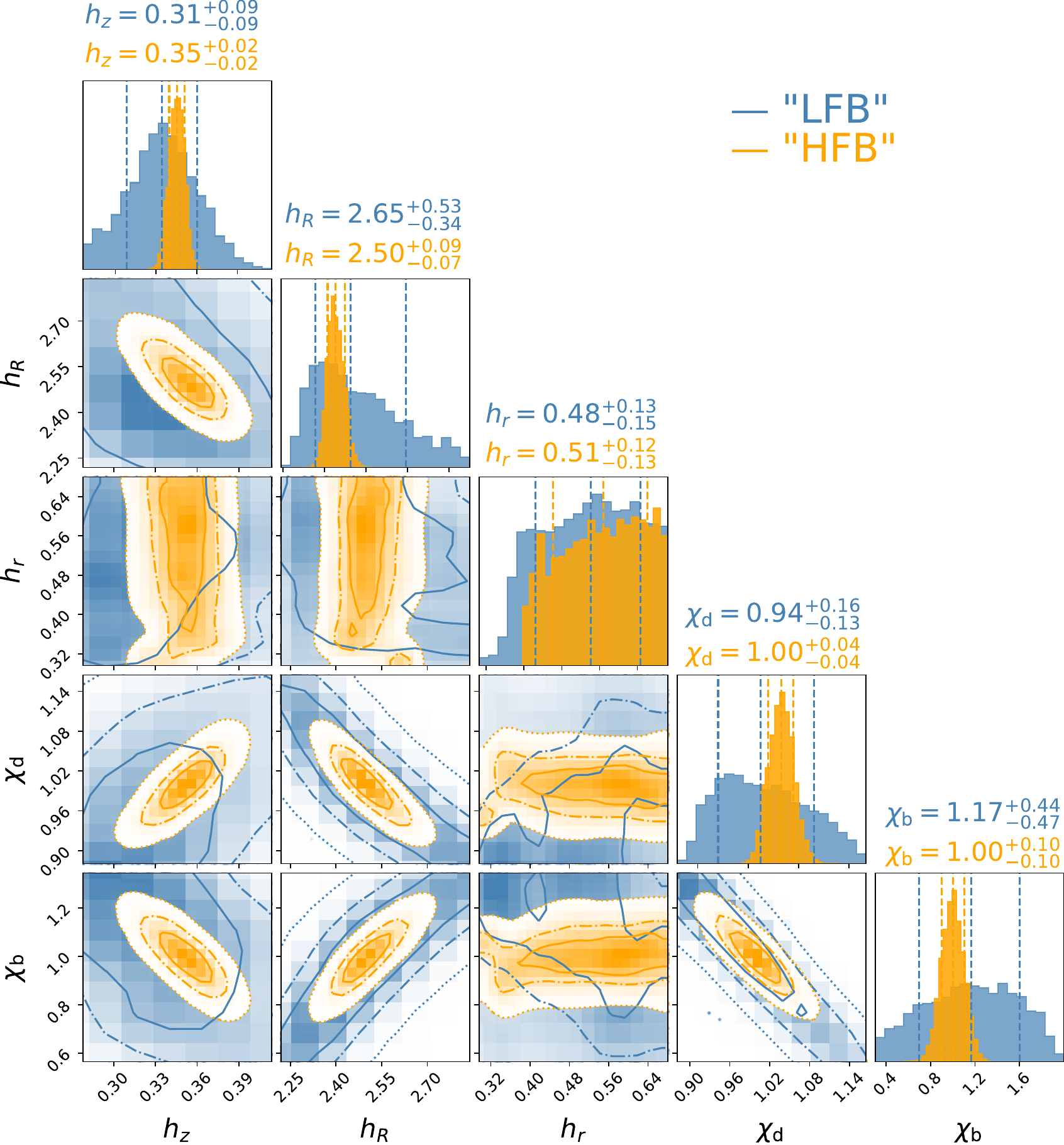}
\caption{
The probability distributions of the Galactic structure model parameters constrained by the YMP resulting from a realization of the Galactic DWDs using the enhanced filter to the ``LFB''  (blue) or the ``HFB'' (orange). The input values of these parameters are $\bm{\Theta_0} = \{0.352$\,kpc, $2.5$\,kpc, $0.5$\,kpc, $1,1\}$. The solid, dash-dotted, and dotted lines in the 2 dimension distribution plots show the $0.5\sigma$, $1\sigma$ and $2\sigma$ contours, i.e., containing $38.3\%$, $68.3\%$, and $95.4\%$ of the samples, respectively. The blue/yellow vertical dashed lines from left to right in each of the one dimensional marginalized probability distribution plots indicate the median and the $1\sigma$ credible intervals. 
}
\label{fig:MCMC_optimal_LISA}
\end{figure*}

We obtain constraints on the model parameters $\bm{\Theta} = \{h_z, h_R, h_r,\chi_{\rm d},\chi_{\rm b}\}$ by using the MCMC method to match the YMP obtained from the simulated GW signals detected by LISA and Taiji with different filters. Figure~\ref{fig:MCMC_optimal_LISA} shows the posterior distribution for one realization for LISA. We show the posterior by applying enhanced filter for ``LFB'' and ``HFB'' in this figure for illustration. Table~\ref{table:constrain results} lists the constraint results by using different filters for different frequency range as well as the whole frequency range for LISA and Taiji. To avoid the bias caused by mock DWD samples, we generate totally 10 realizations for the Galactic DWDs and calculate the constrain results for each realization. We take the average value of the median, upper limit and lower limit of the results from these 10 realizations as the final result, and summarised them in Table~\ref{table:constrain results}. The results from different realizations show slightly difference, therefore the errors mainly come from the parameter constraints. In this table, the results of two different filters (enhanced and top-hat filter) are showed. These filters are applied to different frequency range of sources, they are ``LFB'', ``HFB''. We additionally show the results obtained by using entire frequency range sources in our simulated sample, providing a comparative analysis. It can be seen that when applying the enhanced filter to the ``LFB'', the scale height and scale length for thin disk can be constrained with an accuracy of $0.09$\,kpc ($0.08$\,kpc) and $0.56$\,kpc  ($0.50$\,kpc), the scale radius of bulge can be constrained with an accuracy of $0.17$\,kpc ($0.15$\,kpc) for LISA (Taiji). By applying a top-hat filter to the ``LFB'', we are able to constrain the scale height $h_z$, scale length $h_R$ of thin disk, the scale radius $h_r$ of bulge with an accuracy of $0.11$\,kpc ($0.11$\,kpc), $0.46$\,kpc ($0.39$\,kpc) and $0.34$\,kpc ($0.30$\,kpc) respectively, for LISA (Taiji). The results show a good accuracy for the measurement of Galactic structure by using the technique introduced in this work. And we can also see that the constrain of the scale height for thin disk is significantly better than other structure parameters. It is reasonable because the scale height has a greater impact on the spatial distribution of DWDs, compared to the scale length for disk and the scale radius for bulge. 

We also obtain constraints on the Galactic structure parameters by using the GW spectrum from ``HFB'' and applying the enhanced filter (see the fourth row in Tab.~\ref{table:constrain results}). In this case, $h_z$, $h_R$ and $h_r$ can be constrained with an uncertainty of $0.04$\,kpc ($0.03$\,kpc), $0.13$\,kpc ($0.11$\,kpc) and $0.22$\,kpc ($0.21$\,kpc), respectively, for LISA (Taiji), relatively more accurate than those obtained by using low frequency range sources. Note that using the spatial distribution of the resolved individual DWDs (extract from the observations) to constrain the Galactic structure has been studied well in the literature \citep{Korol2019,Lamberts2019, Li2020, Georgousi2023, Wilhelm2021, Adams2012, Gao2023}, which can give tight constraints on the Galactic structure parameters. The method presented here uses the GW spectrum combined by the GW signals from ``HFB'', i.e. the range dominated by resolved sources. The method that taking consider the sources at high frequency range as a spectrum as a whole, which is different from that using the spatial distribution of individual DWDs extracted from the observations gives constraints on the parameters with similar accuracy.

For the comparison purpose, we also calculate the constrain results from total frequency range ($10^{-4}\sim1.5\times10^{-2}\unit{Hz}$) by applying the enhanced filter (see the last row in Tab.\ref{table:constrain results}). In this case, the constraint results are the tightest. Combined with this situation, we can see from the whole results that the ``HFB'' contributes almost all of the constraint accuracy.

For the top-hat window selection, we set $v_{\rm min} = 7 \times 10^{-4}$ which is the frequency near the intersection of the foreground signal and detector noise as the lower limit for ``LFB''. To validate this choose, we conducted experiments by expanding the frequency range to $6\times10^{-4}\sim1.4\times10^{-3}$ and also shifting it to $8\times10^{-4}\sim1.4\times10^{-3}$. Both scenarios show worse constrain results than previous frequency range, thereby confirming the reasonability of our chosen for the lower limit for ``LFB''.

Apart from three structure parameters for thin disk and bulge, we additionally define the normalization factor $\chi$ for different Galactic structure and these parameters show the uncertainty of the signal amplitude in our estimation. The MCMC results for these normalization parameters are also summarized in Table~\ref{table:constrain results}. It can be seen from the results that the amplitude for thin disk can be constrained better than the bulge. It is expected due to the fact that the GW signal generated from thin disk is stronger than that from the bulge.

We adopt two different length for each time chunk which are 0.5 and 1.5 days, and the constrain results for them show little difference. Therefore, it is reasonable to use $\Delta T = 1\unit{day}$ in our calculations for demonstration purposes.

We also consider the influence of applying varying S/N threshold. Specifically, we set three higher S/N threshold as 10, 15 and 20 and find that there are slightly changes for the constrain results of ``LFB'' and ``HFB'' compared with setting the threshold as 7. Therefore, for the demonstrative purpose, the results shown in this work are based on the S/N threshold of 7.

Here we should note that the pattern function adopted in this work uses the low-frequency approximation \citep{cutler1998}, which may be not applicable for those sources with high frequencies ($\textgreater$10 mHz). When the GW wavelength is smaller than the arm-length of the GW detectors (for LISA/Taiji), the detector’s antenna pattern function has only monopole, quadruple, and sextupole components \citep{Cornish2001, Cornish2003, Kudoh2005, Taruya2005, Taruya2006, Liang2023}. Most of the resolvable mock DWD sources in this work have frequency smaller than 10 mHz (with wavelength larger than the arm length of LISA and Taiji) and only a tiny fraction of the mock DWDs have frequencies larger than $10$\,mHz and the highest frequencies of the mock DWDs are about $15$\,mHz. We check whether the constraining results shown in Figure~\ref{fig:MCMC_optimal_LISA} are affected these high-frequency sources that do not satisfy the low-frequency approximation and find that the constraining results are not affected much by removing those high-frequency DWDs from the HFB. For demonstration purpose, we do not intend to present a separate consideration of these high-frequency sources that do not satisfy the low-frequency approximation but it deserves further study in the future.

\begin{table*}
\centering
\caption{Constraints on the structure parameters of the Galaxy, obtained by using mock observations for LISA and Taiji, generated by setting $(h_z, h_R, h_r, \chi_{\rm d}, \chi_{\rm b})=(0.352$\,kpc,$2.5$\,kpc, $0.5$\,kpc, $1, 1)$.
}
\renewcommand{\arraystretch}{1} 
%
%
\begin{tabular}{ccccccccccc} \hline \hline
%
\multirow{2}{*}{Detector} & \multirow{2}{*}{Filter} &
\multicolumn{9}{c}{Galaxy structure parameters}\\ \cline{3-11}
& & $h_z$(kpc)  & &  $h_R$(kpc)&&$h_r$ (kpc)  & & $\chi_{\rm d}$&&$\chi_{\rm b}$ \\ \cline{1-11} 
%
LISA     & \multirow{2}{*}{enhance (``LFB'')
 }& $0.34^{+0.09}_{-0.09}$  &  &$2.65^{+0.56}_{-0.38}$ &  &$0.46^{+0.15}_{-0.17}$  &   &$0.96^{+0.18}_{-0.15}$ & &$1.17^{+0.48}_{-0.50}$ \\
 Taiji & &$0.33^{+0.08}_{-0.08}$  &  &$2.56^{+0.50}_{-0.30}$ &  &$0.48^{+0.15}_{-0.14}$  &   &$0.98^{+0.14}_{-0.15}$ & &$1.01^{+0.48}_{-0.44}$\\
 \cline{1-11}
 LISA&  \multirow{2}{*}{top-hat (``LFB'')
   }& $0.34^{+0.11}_{-0.11}$ & & $2.64^{+0.46}_{-0.35}$ & & $0.70^{+0.33}_{-0.34}$ &  & $0.94^{+0.20}_{-0.15}$ & & $1.19^{+0.38}_{-0.54}$    \\
 Taiji&   & $0.34^{+0.10}_{-0.11}$ & & $2.61^{+0.39}_{-0.31}$ & & $0.69^{+0.30}_{-0.30}$ &  & $0.95^{+0.15}_{-0.10}$ & & $1.13^{+0.35}_{-0.45}$\\
   \cline{1-11}
 LISA   &  \multirow{2}{*}{enhance (``HFB'')
  }&$0.35^{+0.04}_{-0.03}$ &  &$2.50^{+0.13}_{-0.11}$ &  &$0.47^{+0.22}_{-0.19}$ &  &$1.00^{+0.06}_{-0.06}$ & &$0.99^{+0.15}_{-0.15}$  \\
  Taiji&  &$0.35^{+0.03}_{-0.03}$  &  &$2.50^{+0.11}_{-0.10}$ &  &$0.46^{+0.21}_{-0.19}$  &   &$1.00^{+0.06}_{-0.05}$ & &$1.00^{+0.15}_{-0.14}$\\
   \cline{1-11}
 LISA   &  \multirow{2}{*}{top-hat (``HFB'')
  }&$0.34^{+0.04}_{-0.04}$ &  &$2.54^{+0.13}_{-0.13}$ &  &$0.76^{+0.30}_{-0.35}$ &  &$0.98^{+0.06}_{-0.04}$ & &$1.08^{+0.25}_{-0.25}$  \\
  Taiji&  &$0.34^{+0.03}_{-0.04}$  &  &$2.51^{+0.12}_{-0.11}$ &  &$0.70^{+0.29}_{-0.29}$  &   &$0.99^{+0.05}_{-0.04}$ & &$1.02^{+0.22}_{-0.21}$\\
   \cline{1-11}
 LISA   &  \multirow{2}{*}{enhance$^{*}$
  }&$0.35^{+0.03}_{-0.03}$ &  &$2.50^{+0.12}_{-0.10}$ &  &$0.55^{+0.21}_{-0.19}$ &  &$1.00^{+0.05}_{-0.05}$ & &$0.99^{+0.14}_{-0.13}$  \\
  Taiji&  &$0.35^{+0.03}_{-0.03}$  &  &$2.52^{+0.11}_{-0.10}$ &  &$0.60^{+0.17}_{-0.20}$  &   &$0.99^{+0.04}_{-0.04}$ & &$1.01^{+0.12}_{-0.12}$\\
   \cline{1-11}
\hline \hline
%
\end{tabular}
\begin{tablenotes}
\footnotesize
\item Notes: the first and second columns (from left to right) list the space GW detectors (LISA and Taiji) and two different filters (enhanced filter and top-hat filter) considered in this paper. These filters are applied on three different kind of sources. Except the ``LFB'' and ``HFB'', the ``$\rm enhance^{*}$'' with a superscript represents the results using all frequency range DWDs in our simulated sample and we take this situation as a comparison. The third, fourth, fifth, sixth, and seventh columns list the results on constraining the Galactic structure parameters and the normalized amplitude parameters via the MCMC fitting. The detailed description for the ``top-hat filter'' and the ``optimal filter'' can be seen in Section~\ref{sec:filters}. The frequency window for ``top-hat filter'' is $7\times10^{-4}\sim1.4\times10^{-3}$ and ``$1.4\times10^{-3}\sim1.5\times10^{-2}$'' for ``LFB'' and ``HFB'', respectively. The median, $1\sigma$ upper limit and $1\sigma$ lower limit of constrain results list in this table are derived from the average of the results of the constraints obtained from 10 different realizations of DWD samples. 
\end{tablenotes}
%
%
%
%
\label{table:constrain results}
\end{table*}

\section{Conclusions} 
\label{sec:dis&con}

In this paper, we present an alternative method to constrain the Galactic structure by using the YMP of the anisotropic GW signal from the Galactic DWDs to be detected by space GW detectors.
The anisotropic GW signal is mainly due to the contribution from the DWDs in the thin disks and the spherical bulge. We adopt the BSE models to produce DWD template banks, generate mock samples of the Galactic DWDs for any given set of Galactic structure model parameters, and consequently the GW signals combined from the mock Galactic DWDs in both the time domain and the frequency domain, as well as the YMP of the GW signal in the time domain. The anisotropic distribution of the Galactic DWDs is encoded in the YMP of the GW signal, simply due to the yearly varying pattern functions of the space GW detectors with direction changes relative to the sources. We first consider the phase noise that induced by the randomly distributed phases of DWDs in different time chunks that may affect the YMP and cannot be ignored, and derive analytic expressions for estimating the phase noise. We use the MCMC method to fit the mock YMP of the GW signals detected by LISA/Taiji by considering both the detector's noise and the phase noise, and extract the Galactic structure model parameters. We find that the input Galactic structure model parameters (for the thin disk and the bulge) can be well reconstructed with quite high accuracy, the scale height and length of the Galactic thin disk and the scale radius of bulge can be constrained to a fractional accuracy
 $\sim$ $30\%$, $30\%$, $40\%$ (or $20\%$, $10\%$, $40\%$) by using the low-frequency (or high-frequency) signal detected by LISA or Taiji, which validates the new method presented in this paper. In comparison to other works in the literature utilizing the GW foreground in the frequency domain \citep[e.g.,][]{Benacquista2006, Breivik2020} for constraining Galactic structure, our approach provides an alternative methodology that is not only more direct but also computationally efficient, while achieving comparable or even higher accuracy.

We note here that the measured time-domain GW signals from the Galactic DWDs detected by space GW detectors is directly applied in our method. We also further consider the distinction of the signal in two different bands, the ``LFB'' and the ``HFB'', with the former contributing by un-resolvable DWDs radiating GWs at lower frequencies and the latter one mainly contributing by resolvable DWDs radiating GWs at higher frequencies. The ``LFB'' and the ``HFB'' signals are separately used to obtain constraints on the Galactic structure parameters. We find that the constraints obtained by using the ``HFB'' band signal are stronger than those by using the ``LFB'' signal. We also introduce some filters for both the ``LFB'' and ``HFB'' signals and find choosing appropriate ``enhanced'' filters can lead to tighter constraints. 
Our analysis mainly relies on the modulation of the GW signal, even for the “HFB”. The previous analyses of resolved DWDs went a step further to constrain the Galactic structure by directly using the sky distribution of these sources with GW inferred locations \citep[e.g.,][]{Korol2019, Lamberts2019, Georgousi2023, Wilhelm2021, Adams2012, Gao2023}. It is interesting to mention that our approach in this paper by just using the bulk modulation of the resolved sources leads to similar constraints on the Galaxy shape, which suggests that our approach provides a different but also efficient way of using the GW signal to constrain the shape of the Galaxy.

\section*{acknowledgments}
This work is partly supported  by the National Key Research and Development Program of China (grant no. 2020YFC2201400, 2022YFC2205201, 2021YFC2203003), the National Natural Science Foundation of China (grant nos. 12273050, 11991052), and the Strategic Priority Research Program of the Chinese Academy of Sciences (grant no. XDB0550300).
\section*{Data Availability}
The data underlying this article will be shared on reasonable request to the corresponding author.

%



\software{astropy \citep{2013A&A...558A..33A,2018AJ....156..123A}
          }


\appendix

\section{Derivation of noise for the YMP and the detector}
\label{sec:appendixA}

In this appendix, we provide a detailed derivation of the noise from the yearly modulate profile (abbreviated as ``YMP'') and the detector. We define the average of the strain in the $k$-th time chunk as $S_k$, which contains both the signal and the detector noise, i.e., 
\begin{equation}
S_k=\frac{1}{\Delta{T}} \int_{k\Delta T}^{(k+1)\Delta T} (h(t)+n(t))^2dt.
\end{equation}
Here the combined GW signal of all DWDs is denoted as $h(t)$, the detector noise is denoted as $n(t)$, and the length of each time chunk for average is $\Delta T$. We can obtain the variance of $S_k$ by calculating the ensemble average $\left< S_k S_{k'}\right>$ and $\left<S_k\right>\left<S_{k'}\right>$, where the subscripts $k$ and $k'$ denote two time chunks. For simplicity, we use $\int_k$ and $\int_{k'}$ to denote the integration over the time chunk $k$ and $k'$, respectively. The ensemble average for $S_k S_{k'}$ can be written as
\begin{equation}
\left<S_kS_{k'}\right>=\frac{1}{\Delta T^2} \left<\int_k(h(t)+n(t))^2dt \int_{k'}(h(t)+n(t))^2dt\right>.
\end{equation}

Due to the Gaussian and stationary detector noise, as well as the non-correlation between the signal and detector noise, the above Equation can be rewritten as
\begin{multline}
\left<S_kS_{k'}\right>=\frac{1}{\Delta T^2}\Bigg( \left<\int_k h^2(t)dt \int_{k'}h^2(t)dt\right>+\left<\int_k h^2(t)dt\right>\left<\int_{k'} n^2(t)dt\right>
+\\
\left<\int_{k'} h^2(t)dt\right>\left<\int_k n^2(t)dt\right>+\left<\int_k n^2(t)dt\int_{k'} n^2(t)dt\right> \Bigg).
\end{multline}
Similarly, we can get 
\begin{multline}
\left<S_k\right>\left<S_{k'}\right>=\frac{1}{\Delta T^2}\Bigg( \left<\int_k h^2(t)dt\right>\left<\int_{k'}h^2(t)dt\right>+\left<\int_{k}h^2(t)dt\right>\left<\int_{k'} n^2(t)dt\right>\\
+\left<\int_{k'} h^2(t)dt\right>\left<\int_k n^2(t)dt\right>+\left<\int_kn^2(t)dt\right>\left<\int_{k'} n^2(t)dt\right> \Bigg).
\end{multline}
Therefore, the variance of $S_k$ can be written as
\begin{equation}
\sigma_{k,k'}^2 = \left<S_kS_{k'}\right>-\left<S_k\right>\left<S_{k'}\right>= \left<F_kF_{k'}\right>-\left<F_k\right>\left<F_{k'}\right>+\left<N_kN_{k'}\right>-\left<N_k\right>\left<N_{k'}\right>
\label{sigmaS}
\end{equation}
where we denote
\begin{equation}
F_k = \frac{1}{\Delta T}\int_k h^2(t) dt ,\qquad
N_k = \frac{1}{\Delta T}\int_k n^2(t) dt 
\label{FandN}
\end{equation}
for the $k$-th time chunk.

The GW signal $h(t)$ at any given time $t$ is given by 
\begin{equation}
h(t) = \sum_n( A_n\cos(2\pi \nu_n t+\phi_n)+B_n\sin(2\pi \nu_nt+\phi_n) ),
\label{eq:A_ht}
\end{equation}
where the summation is over all DWDs, $A_n$ and $B_n$ are the constant coefficient for ``+'' and ``$\times$'' polarization for the $n$-th DWD, respectively, $\phi_n$ is the phase of the strain signal from the $n$-th DWD. The Doppler term $\phi_{\rm D}$ in Equation~\eqref{eq:phi_D} is omitted in our derivation for the reason that the fractional changes due to this term on the YMP is smaller than $1\%$ when the average length $\Delta T = 1\unit{day}$ is adopted. Since the variation of the detector antenna in each time chunk can be approximated as constant, therefore the terms $F^+$ and $F^{\times}$ in each time chunk can be written in constant coefficient $A_n$ and $B_n$ respectively. Then, we have
\begin{multline}
\begin{aligned}
h^2(t)
&= \sum_{n,m} (A_n\cos(2\pi \nu_n t+\phi_n)+B_n\sin(2\pi \nu_nt+\phi_n)) \times(A_m\cos(2\pi \nu_m t+\phi_m)+B_m\sin(2\pi \nu_mt+\phi_m))\\
&=\frac{1}{2}\sum_{n,m}A_nA_m\cos(2\pi(\nu_n+\nu_m)t+\phi_n+\phi_m)+\frac{1}{2}\sum_{n,m}A_nA_m\cos(2\pi(\nu_n-\nu_m)t+\phi_n-\phi_m)\\
&-\left[\frac{1}{2}\sum_{n,m}B_nB_m\cos(2\pi(\nu_n+\nu_m)t+\phi_n+\phi_m)-\frac{1}{2}\sum_{n,m}B_nB_m\cos(2\pi(\nu_n-\nu_m)t+\phi_n-\phi_m)\right]\\
&+2\left[\frac{1}{2}\sum_{n,m}A_n B_m\sin(2\pi(\nu_n+\nu_m)t+\phi_n+\phi_m)-\frac{1}{2}\sum_{n,m}A_nB_m\sin(2\pi(\nu_n-\nu_m)t+\phi_n-\phi_m)\right].
\label{hf2}
\end{aligned}
\end{multline}
In order to calculate the $F_k$ which is defined in Equation~\eqref{FandN}, we need to integrate Equation~\eqref{hf2}. To make the derivation looks clearer, we define the integration for the first two terms of Equation~\eqref{hf2} as $F_{1,k}$ (the terms contain $A_nA_m$ and the cosine), and show the calculation process as
\begin{multline}
\begin{aligned}
F_{1,k}&= \sum_{n,m}\frac{A_n A_m}{4\pi \Delta T(\nu_n+\nu_m)}( \sin(2\pi(\nu_n+\nu_m)(k+1)\Delta T +\phi_n+\phi_m) - \sin(2\pi(\nu_n+\nu_m)k\Delta T +\phi_n+\phi_m) )\\
&+ \sum_{n\neq m}\frac{A_n A_m}{4\pi \Delta T(\nu_n-\nu_m)}( \sin(2\pi(\nu_n-\nu_m)(k+1)\Delta T+\phi_n-\phi_m) - \sin(2\pi(\nu_n-\nu_m)k\Delta T+\phi_n-\phi_m) ) + \sum_{n}\frac{A_n^2}{2}\\
&= \sum_{n,m}\frac{A_n A_m}{2\pi \Delta T(\nu_n+\nu_m)}\cos(\pi(\nu_n+\nu_m)(2k+1)\Delta T+\phi_n+\phi_m)\sin(\pi(\nu_n+\nu_m)\Delta T) \\
&+ \sum_{n\neq m}\frac{A_n A_m}{2\pi \Delta T(\nu_n-\nu_m)} \cos(\pi(\nu_n-\nu_m)(2k+1)\Delta T+\phi_n-\phi_m)\sin(\pi(\nu_n-\nu_m)\Delta T)+ \sum_{n}\frac{A_n^2}{2}\\
&= \sum_{n,m}\frac{A_n A_m}{2}\cos(\pi(\nu_n+\nu_m)(2k+1)\Delta T +\phi_n+\phi_m){\rm sinc}(\pi(\nu_n+\nu_m)\Delta T) \\
&+ \sum_{n, m}\frac{A_n A_m}{2} \cos(\pi(\nu_n-\nu_m)(2k+1)\Delta T+\phi_n-\phi_m){\rm sinc}(\pi(\nu_n-\nu_m)\Delta T).
\label{F1k}
\end{aligned}
\end{multline}
With the typical frequency range of DWD and the 1 day length for $\Delta T$, one can easily get $\Delta T (\nu_n+\nu_m) \gg 1$, then according to the property of the the \sinc function, Equation~\eqref{F1k} can be approximated to
\begin{equation}
F_{1,k}
\approx \sum_{n, m}\frac{A_n A_m}{2} \cos(\pi(\nu_n-\nu_m)(2k+1)\Delta T+\phi_n-\phi_m){\rm sinc}(\pi(\nu_n-\nu_m)\Delta T).
\end{equation}
Similarly, we can calculate the integral of other terms in Equation~\eqref{hf2}. The final results can be summarised as 
\begin{multline}
F_k\approx \sum_{n, m}(\frac{A_n A_m+B_n B_m}{2}) \cos(\pi(\nu_n-\nu_m)(2k+1)\Delta T+\phi_n-\phi_m){\rm sinc}(\pi(\nu_n-\nu_m)\Delta T)\\
-\sum_{n, m}A_nB_m\sin(\pi(\nu_n-\nu_m)(2k+1)\Delta T+\phi_n-\phi_m){\rm sinc}(\pi(\nu_n-\nu_m)\Delta T).
\end{multline}
Furthermore, we calculate
\begin{multline}
\begin{aligned}
F_{1,k} F_{1,k'}\approx
\sum_{n, n', m, m'}
&\frac{A_nA_mA_{n'}A_{m'}}{4} \cos(\pi(\nu_n-\nu_m)(2k+1)\Delta T+\phi_n-\phi_m) {\rm sinc}(\pi(\nu_n-\nu_m)\Delta T)\\
&\cos(\pi(\nu_{n'}-\nu_{m'})(2k'+1)\Delta T+\phi_{n'}-\phi_{m'}) {\rm sinc}(\pi(\nu_{n'}-\nu_{m'})\Delta T)\\
\approx
\sum_{n, n', m, m'}
&\frac{A_nA_mA_{n'}A_{m'}}{8} {\rm sinc}(\pi(\nu_n-\nu_m)\Delta T){\rm sinc}(\pi(\nu_{n'}-\nu_{m'})\Delta T)\\
&\cos(\pi((2k+1)(\nu_n-\nu_m)+(2k'+1)(\nu_{n'}-\nu_{m'}))\Delta T+\phi_n-\phi_m+\phi_{n'}-\phi_{m'}) \\
+
\sum_{n, n', m, m'}
&\frac{A_nA_mA_{n'}A_{m'}}{8} {\rm sinc}(\pi(\nu_n-\nu_m)\Delta T){\rm sinc}(\pi(\nu_{n'}-\nu_{m'})\Delta T)\\
&\cos(\pi((2k+1)(\nu_n-\nu_m)-(2k'+1)(\nu_{n'}-\nu_{m'})\Delta T+\phi_n-\phi_m-\phi_{n'}+\phi_{m'}).
\label{F0Fs}
\end{aligned}
\end{multline}
Similarly, $F_{1,k} F_{1,k'}$ is one of the nine terms produced by multiplying $F_{k}$ by $F_{k'}$. We show the detailed calculation for this term rather than all nine terms to make the derivation looks clearly. If we take 
$\phi_n-\phi_m+\phi_{n'}-\phi_{m'}=0$,
the first term in Equation~\eqref{F0Fs} has a significant value. Due to these phases are independently random number, we can obtain the relations $(n=m, n'=m')$ or $(n=m', n'=m)$. The second term can be approximated in the same way. Therefore,
\begin{equation}
\langle F_{1,k} F_{1,k'} \rangle\approx
\sum_{n, m}\frac{A_n^2A_m^2}{4} + \sum_{n\neq m} \frac{A_n^2A_m^2}{4} {\rm sinc}^2(\pi(\nu_n-\nu_m)\Delta T)\cos(2\pi (k'-k)\Delta T(\nu_n-\nu_m)),
\end{equation}
where $k \leq k'$. For the other 8 terms $F_{1,k}F_{2,k'}$, $F_{1,k}F_{3,k'}$, $F_{2,k}F_{1,k'}$, $F_{2,k}F_{2,k'}$, $F_{2,k}F_{3,k'}$, $F_{3,k}F_{1,k'}$,$F_{3,k}F_{2,k'}$, $F_{3,k}F_{3,k'}$, we can similarly get the results of ensemble average. We summary the finally results as
\begin{multline}
\begin{aligned}
&\langle F_k F_{k'} \rangle\approx \sum_{n,m}\frac{(A_n^2+B_n^2)(A_m^2+B_m^2)}{4}+\sum_{n\neq m}\frac{(A_n^2+B_n^2)(A_m^2+B_m^2)}{4}{\rm sinc}^2(\pi(\nu_n-\nu_m)\Delta T)\cos(2\pi(k'-k)\Delta T(\nu_n-\nu_m)).
\end{aligned}
\end{multline}
Similarly, one can easily obtain
\begin{equation}
\left<F_k\right>\left<F_{k'}\right>=\sum_{n,m}\frac{A_n^2A_m^2}{4}+\sum_{n,m}\frac{B_n^2B_m^2}{4}+\sum_{n, m}\frac{2A_nA_mB_nB_m}{4},
\end{equation}
and
\begin{equation}
\left<F_k\right>=\frac{\sum_{n} {A_n^2}+\sum_{n}{B_n^2}}{2},
\label{Fk}
\end{equation}
therefore, the signal terms (the first two terms in Equation~\eqref{sigmaS}) for the variance of $S_k$ can be written as
\begin{multline}
\begin{aligned}
&\langle F_k F_{k'} \rangle  -\left<F_k\right>\left<F_{k'}\right> \approx \sum_{n\neq m}\frac{(A_n^2+B_n^2)(A_m^2+B_m^2)}{4}{\rm sinc^2}(\pi(\nu_n-\nu_m)\Delta T)\cos(2\pi(k'-k)\Delta T(\nu_n-\nu_m)).
\label{FF-F_F}
\end{aligned}
\end{multline}
Assuming a high number density of sources within the frequency range of interest, which is generally valid for the huge number of DWD sources, the summation can be reformulated as an integral
\begin{equation}
\langle F_kF_{k'}\rangle-\langle F_k\rangle \langle F_{k'}\rangle 
\approx \int_{0}^{+\infty} (S_A(\nu_1) +S_B(\nu_1))(S_A(\nu_2)+S_B(\nu_2)){\rm sinc}^2(\pi(\nu_1 - \nu_2)\Delta T)e^{-i2\pi (k'-k)\Delta T(\nu_1-\nu_2) }d\nu_1d\nu_2.
\label{eq:From discrete to continuous }
\end{equation}
The sine term from the first term of Equation~\eqref{FF-F_F} after using Euler's formula is omitted because the $\rm{sinc^2}$ is symmetric, therefore the integral for $\rm{sinc^2}$ times sine is 0. We define $S_A(\nu)$ as half of the summation of the squares of the amplitudes ``+'' polarization from all the sources in per unit frequency interval. $S_B(\nu)$ is defined similar with $S_A(\nu)$ but for ``$\times$'' polarization. Because the integral is non-zero only if $\nu_1$ and $\nu_2$ are very close, we define $\nu_1 -\nu_2 =\delta$ which $\delta$ has very small value. Therefore, we can perform a variable substitution and rewrite Equation~\eqref{eq:From discrete to continuous } in the form of variable separation as
\begin{eqnarray}
\left<F_kF_{k'}\right>-\left<F_k\right>\left<F_{k'}\right> 
&\approx&\int_{0}^{+\infty} (S_A(\nu_2)+S_B(\nu_2))^2d\nu_2\int_{-\infty}^{+\infty} {\rm sinc}^2(\pi\delta \Delta T)e^{-i2\pi (k'-k)\Delta T\delta }d\delta
\end{eqnarray}
According to the mathematical manual, the fft result for $\rm{sinc^2}$ are:
\begin{equation}
\int_{-\infty}^{+\infty} \rm{sinc^2(\pi ax)}e^{-i2\pi \xi x}dx=\frac{1}{|a|}tri(\frac{\xi}{a}),
\label{tri}
\end{equation}
we can rewrite the formula as
\begin{multline}
\begin{aligned}
&\left<F_kF_{k'}\right>-\left<F_k\right>\left<F_{k'}\right> \approx \frac{1}{\Delta T}{\rm{tri}}(k'-k)\int_{0}^{+\infty} (S_A(\nu)+S_B(\nu))^2d\nu=\frac{1}{\Delta T}\delta_{k',k}\int_0^{+\infty}S^2_s(\nu)d\nu. 
\end{aligned}
\end{multline}
Where the $\delta_{k',k}$ is the Kronecker delta, which means the covariance between different time chunks can be ignored. Therefore, in our calculation, we only consider the variance of each time chunk. We here define $S_s(\nu)=S_A(\nu)+S_B(\nu)$ as the total power spectral density of all sources which contains resolved and unresolved sources. Finally, according to this equation, one can estimate the fluctuation caused by the phases of DWD sources with given spectrum. 

After deriving the expression for the signal part of $\sigma^2$, we calculate the detector noise term of it. According to Equation~\eqref{sigmaS}, this term can be written as:
\begin{equation}
\frac{1}{\Delta T^2}\left<\int_k n^2(t)dt \int_{k'} n^2(t) dt \right> -\frac{1}{\Delta T^2}\left<\int_k n^2(t) dt \right>\left<\int_{k'} n^2(t) dt \right>
\label{noise_sigma}
\end{equation}
According to the inverse Fourier transform and the properties of Dirac delta function, we have
\begin{multline}
\begin{aligned}
&\left<\int_k n^2(t)dt\int_{k'} n^2(t') dt' \right>
=\int_k\int_{k'} \left( \int_{-\infty}^{+\infty}  \left< n(\nu_1)n(\nu_2)n(\nu_3)n(\nu_4) \right> e^{i2\pi t(\nu_1+\nu_2)}e^{i2\pi t'(\nu_3+\nu_4)}d\nu_1d\nu_2d\nu_3d\nu_4 \right) dt dt'\\
=&\frac{1}{2}\int_k\int_{k'} \left(\int_{-\infty}^{+\infty} S_{\rm n}(\nu_1) S_{\rm n}(\nu_2)e^{i2\pi t(\nu_1+\nu_2)}e^{-i2\pi t'(\nu_1+\nu_2)} d\nu_1 d\nu_2\right) dtdt'
+\frac{1}{4}\int_k\int_{k'} \left( \int_{-\infty}^{+\infty} S_{\rm n}(\nu_1) S_{\rm n}(\nu_3) d\nu_1 d\nu_3\right)dt dt'.
\label{N1term}
\end{aligned}
\end{multline}
The second step is obtained by using the fact 
\begin{equation}
\langle n(\nu_1) n(\nu_2) n(\nu_3) n(\nu_4)\rangle=\langle n(\nu_1) n(\nu_2) \rangle\langle n(\nu_3) n(\nu_4) \rangle + \langle n(\nu_1) n(\nu_3) \rangle\langle n(\nu_2) n(\nu_4) \rangle + \langle n(\nu_1) n(\nu_4) \rangle\langle n(\nu_2) n(\nu_3) \rangle
\end{equation}
and
\begin{equation}
\langle n(\nu_1) n(\nu_2)\rangle=\langle n(\nu_1) n^*(-\nu_2)\rangle = \frac{1}{2}S_{\rm n}(\nu_1)\delta(\nu_1+\nu_2),
\end{equation}
where the $S_{\rm n}(\nu)$ is the one-side noise power spectrum. Note that the $S_{\rm n}(\nu)$ still has value when frequency lower than 0 in Equation~\eqref{N1term} because the integral for $n(\nu)$ is from $-\infty$ to $+\infty$. Therefore, there has an extra factor 2 for each integral when changing the lower limit of integral from $-\infty$ to $0$. Subsequently, the second term of Equation~\eqref{N1term} can be further simplified as 
%
%
$\Delta T^2\bigg(\int_{0}^{+\infty} S_{\rm n}(\nu) d\nu\bigg)^2$.
%
%
Then, Equation~\ref{N1term} can be written as
\begin{multline}
\left<\int_k n^2(t) dt\int_{k'} n^2(t) dt \right> = \frac{1}{2}\int_k\int_{k'} \left(\int_{-\infty}^{+\infty} S_{\rm n}(\nu_1) S_{\rm n}(\nu_2)e^{i2\pi t(\nu_1+\nu_2)}e^{-i2\pi t'(\nu_1+\nu_2)} d\nu_1 d\nu_2\right) dtdt' +\Delta T^2\bigg(\int_{0}^{+\infty} S_{\rm n}(\nu) d\nu\bigg)^2\\
= \frac{\Delta T^2}{2}\int_{-\infty}^{+\infty} S_{\rm n}(\nu_1)S_{\rm n}(\nu_2) {\rm{sinc}}^2(\pi(\nu_1+\nu_2)\Delta T)e^{-i2\pi(\nu_1+\nu_2)(k'-k)\Delta T}d\nu_1d\nu_2 + \Delta T^2\bigg(\int_{0}^{+\infty} S_{\rm n}(\nu) d\nu\bigg)^2\\
=\Delta T^2\int_{-\infty}^{+\infty} S_{\rm n}(x+y)S_{\rm n}(x-y){{\sinc}^2}(2\pi x\Delta T)e^{-i2\pi2x(k'-k)\Delta T}dxdy + \Delta T^2\bigg(\int_{0}^{+\infty} S_{\rm n}(\nu) d\nu\bigg)^2.
\end{multline}
The third step is obtained by using variable substitution between $v_1,v_2$ and new variables $x,y$ with relationship as $v_1+v_2=2x, v_1-v_2=2y$. Due to the property of the sinc function, only if $x\leq 1/\Delta T$ can make this integrate have large value. Meanwhile, for a such small x, the changes of the power spectral density of the detector can be ignored, therefore, this equation can be written as
\begin{multline}
\left<\int_k n^2(t)dt\int_{k'} n^2(t) dt \right> \simeq \Delta T^2\int_{-\infty}^{+\infty} S_{\rm n}(y)S_{\rm n}(-y)dy \int_{-\infty}^{+\infty} {\rm{sinc}^2}(2\pi x \Delta T)e^{-i2\pi2x(k'-k)\Delta T}dx+ \Delta T^2\left(\int_{0}^{+\infty} S_{\rm n}(\nu) d\nu\right)^2\\
=\Delta T \int_{0}^{+\infty} S_{\rm n}^2(\nu) d\nu {\rm{tri}}(k'-k)+ \Delta T^2\left(\int_{0}^{+\infty} S_{\rm n}(\nu) d\nu\right)^2.
\end{multline}
The second step is obtained by using the formula which is shown in Equation~\eqref{tri}.\par
Similarly, we can calculate the second term for Equation~\eqref{noise_sigma} as
\begin{multline}
\left<\int_k n^2(t) dt \right> = \int_k \left(\int_{-\infty}^{+\infty}  \left< n(\nu_1) n(\nu_2)\right> e^{i 2\pi t (\nu_1+\nu_2)} d\nu_1 d\nu_2\right) dt\\
=\frac{1}{2}\int_k\left(\int_{-\infty}^{+\infty} S_{\rm n}(\nu_1)\delta(\nu_1+\nu_2) e^{2\pi i (\nu_1+\nu_2) t}d\nu_1d\nu_2\right)dt=\Delta T\int_{0}^{+\infty} S_{\rm n}(\nu)d\nu
\end{multline}
Finally, we get:
\begin{equation}
\frac{1}{\Delta T^2}\left<\int_k n^2(t)dt \int_{k'} n^2(t) dt \right>-\frac{1}{\Delta T^2}\left<\int_k n^2(t) dt \right>\left<\int_{k'}n^2(t)dt\right> = \frac{1}{\Delta T} \delta_{k',k} \int_{0}^{+\infty} S_{\rm n}^2(\nu)d\nu
\end{equation}
The final results for $\sigma_k^2$ can be summarized as
\begin{equation}
\sigma_k^2=\frac{1}{\Delta T}\delta_{k',k}\int_{0}^{+\infty} (S_s^2(\nu)+S_{\rm n}^2(\nu)) d\nu,
\end{equation}
which contains the detector noise and the noise caused by signal.
%

\section{Appendix B } 
\label{sec:appendixB}

There is a derivation for the enhanced filter. We denote $S_s(\nu)=S_A(\nu)+S_B(\nu)$ as the total power spectral density of all sources which contains resolved and unresolved sources. It is worth to mention that the derivation below is shown in one time chunk for demonstrate, we omit the subscript ``k'' for simplify. Then we have 
\begin{equation}
\sigma^2 =\frac{1}{\Delta T} \int_{0}^{+\infty}(S_s^2(\nu)+S_{\rm n}^2(\nu))d\nu,\qquad
\langle F\rangle = \int_{0}^{+\infty} S_{\rm{s}}(\nu) d\nu.
\end{equation}
If we filter the time-domain data, which is equivalent to multiply a weighting $w_0(\nu)$ on $S_s(\nu)$ and $S_{\rm n}(\nu)$, we get
\begin{equation}
\label{eq:sigma_w and F_ur_w}
\sigma_w^2 = \frac{1}{\Delta T}\int_{0}^{+\infty} w_0^2(\nu)(S_s^2(\nu) + S_{\rm n}^2(\nu))d\nu,\qquad
\langle F_{w} \rangle = \int_{0}^{+\infty} w_0(\nu) S_{\rm{s}}(\nu) d\nu.
\end{equation}
To obtain the filter that effectively enhances the region where we interest, one can minimize the value of $r=\sigma_w^2/\langle F_{w} \rangle^2$ which is the value of noise to signal. To achieve this, we vary the $w$ by $w = w_{\rm{ehc}} + \delta w$, and only keep the first order terms as
\begin{multline}
r_0 + \delta r
\approx \frac{1}{\Delta T}\frac{\int_{0}^{+\infty} w_{\rm{ehc}}^2(1 + 2\delta w/w_{\rm{ehc}})(S^2_s + S^2_n)d\nu}{(\int_{0}^{+\infty} w_{\rm{ehc}} S_s d\nu)^2(1 + \frac{2\int_{0}^{+\infty}\delta w S_sd\nu}{\int_{0}^{+\infty} w_{\rm{ehc}} S_sd\nu})}\\
\approx r_0 + \frac{1}{\Delta T}\frac{2\int_{0}^{+\infty} w_{\rm{ehc}} \delta w(S^2_s + S^2_n)d\nu}{(\int_{0}^{+\infty} w_{\rm{ehc}} S_s d\nu)^2} - \frac{1}{\Delta T }\frac{\int_{0}^{+\infty} w_{\rm{ehc}}^2(S^2_s + S^2_n)d\nu}{(\int_{0}^{+\infty} w_{\rm{ehc}} S_s d\nu)^2 }\frac{2\int_{0}^{+\infty} \delta w S_sd\nu}{\int_{0}^{+\infty} w_{\rm{ehc}} S_sd\nu}.
\end{multline}
We require $\delta r = 0$, then we have
\begin{equation}
\label{eq:deltar=0}
\frac{\int_{0}^{+\infty} w_{\rm{ehc}} \delta w(S^2_s + S^2_n)d\nu}{\int_{0}^{+\infty} w_{\rm{ehc}}^2(S^2_s + S^2_n)d\nu} = \frac{\int_{0}^{+\infty} \delta w S_sd\nu}{\int_{0}^{+\infty} w_{\rm{ehc}} S_sd\nu}.
\end{equation}
If we set the filter as
\begin{equation}
w_{\rm{ehc}}=\frac{S_s}{S_s^2(\nu)+S_{\rm n}^2(\nu)},
\end{equation}
Equation~\eqref{eq:deltar=0} can be satisfied. At that point, $r$ will reach an extremism. Therefore, the weighted value for noise and signal by using the enhanced filter $w_{\rm{ehc}}$ can be written as
\begin{equation}
\sigma_w^2 = \frac{1}{\Delta T}\int_{0}^{+\infty} w_{\rm{ehc}}^2(\nu)(S_s^2(\nu) + S_{\rm n}^2(\nu))d\nu,\qquad
\langle F_{w} \rangle = \int_{0}^{+\infty} w_{\rm{ehc}}(\nu) S_s(\nu) d\nu.
\end{equation}

\bibliography{sample631}{}
\bibliographystyle{aasjournal}



\end{document}